\documentclass[aps,prb,twocolumn,groupedaddress,floats,showpacs]{revtex4}
\usepackage{bm}% bold math

\usepackage{amssymb,amsmath}
\usepackage{latexsym}
\usepackage{graphicx}
\usepackage{epsfig}

\begin{document}

\preprint{APS/123-QED}

\title{Femtosecond Control of the  Magnetization in Ferromagnetic
Semiconductors} 

\author{J. Chovan and 
I. E.  Perakis}

\affiliation{Institute of Electronic Structure \& Laser, Foundation
for Research and Technology-Hellas
and Department of Physics, University of Crete, Heraklion, Greece}

\date{\today}

\begin{abstract}

We develop a theory of 
 collective spin dynamics triggered by ultrafast optical 
excitation of  ferromagnetic  semiconductors.  
Using the density matrix equations of motion
in the mean field approximation and including magnetic anisotropy 
and hole spin dephasing effects,
we predict the development of a 
light--induced magnetization tilt 
during ultra--short time intervals comparable 
to the pulse duration. This femtosecond dynamics  
in the coherent temporal regime 
is governed 
by the interband nonlinear optical polarizations
and is followed by a second temporal regime governed 
by the magnetic anisotropy of the Fermi sea. 
We interpret our numerical results by deriving 
a 
Landau--Gilbert--like  
equation for the collective  spin, which demonstrates 
an ultrafast
correction to the magnetic anisotropy effective 
field 
due to second order 
coherent nonlinear optical processes.
Using the Lindblad semigroup method, 
we also derive  a contribution to the interband  polarization dephasing 
determined by the  Mn spin and the hole spin dephasing. 
Our predicted magnetization tilt 
and subsequent nonlinear dynamics due to the magnetic anisotropy 
can be controlled by varying the optical pulse intensity, 
duration, and helicity 
and can be observed  with  pump--probe 
magneto--optical spectroscopy.
\end{abstract}

\pacs{78.47.J-, 78.20.Ls, 78.30.Fs, 42.50.Md }
\maketitle

\section{Introduction}
\label{sec:intro}

The interaction between itinerant carrier spins and 
localized magnetic moments 
leads to carrier--mediated ferromagnetic order 
 in a wide variety  of systems, ranging 
from ferromagnetic semiconductors such as EuO, EuS, 
chrome spinels, or pyrochlore \cite{nagaev-rev} to 
 manganese oxides (manganites)
 \cite{tokura}
and   III-Mn-V ferromagnetic 
semiconductors  \cite{ohno,mean,review}.
Such materials offer potential 
for novel  spintronics applications \cite{wolf}.
Their
magnetic and 
transport properties are intimately related  and 
can be controlled by varying  
e.g. the  carrier density, spin, and distribution. 

The non--equilibrium and dynamical properties
of ferromagnetic semiconductors and magnetically--ordered systems are
currently under investigation. 
For 
THz spintronic and magnetic devices, 
ultrafast information storage, recovery, and 
processing is required,  
e.g. the development of devices with 
sub-picosecond readout times of the magnetic states.
This goal requires 
femtosecond 
spin manipulation and control. 
The physical processes that govern 
the magnetization dynamics during timescales 
shorter than the characteristic response times 
of the magnetic system
are still under debate. 
During such timescales, the validity of 
conventional thermodynamic 
concepts for describing a magnetic system 
become questionable. 
Ultrafast pump-probe magneto-optical spectroscopy 
can shed light 
into this fundamental problem.
In these experiments, an ultra--short pump optical pulse 
excites 
optical polarizations, non--thermal 
populations, and carrier spins, which then 
trigger a magnetization dynamics
measured  as function of time
(Faraday or Kerr rotation)  \cite{wang-rev}.

To interpret such experiments,
it is useful to distinguish between  different 
stages of time evolution 
of the photoexcited system.
During the initial sub-picosecond regime, 
shorter than the dephasing times 
 or  optical pulse duration,  
the dynamics of the collective magnetization is triggered by  
optical polarizations
and photoexcited carrier 
spins. 
The response of the magnetic system
is   controlled by the nonlinear optical excitation
and conventional 
thermodynamic concepts do not apply.
In this initial regime, magnetic anisotropy can play a role 
by affecting  the 
photoexcited carrier spin.
After the pulse is gone, 
the 
light--induced 
quantum mechanical coherences 
decay  and the photoexcited carriers
relax by interacting with the 
hole Fermi sea. 
The carrier temperature is thus elevated
above
the lattice temperature
within hundreds of femtoseconds.
The hole population is eventually  described by a hot Fermi--Dirac 
distribution, which transfers its 
excess energy 
to the lattice within a few picoseconds.
At the same time, the magnetic axes 
of the system change, due to e.g. the transient temperature elevation. 
Such quasi--thermal changes in the magnetic anisotropy are due to the 
Fermi sea carriers and should be contrasted to the 
magnetic anisotropy contribution of the non--thermal photoexcited carriers 
in the initial temporal regime. 
After the equilibration of the carrier and lattice systems, 
their common temperature 
relaxes 
via a slow (nanosecond) thermal diffusion process, 
which returns the magnetic system to its 
equilibrium configuration. In this paper we discuss 
a mechanism for  coherent non--thermal 
magnetization manipulation 
and neglect all thermal effects due to  elevated carrier 
and spin temperatures.

Most of the ultrafast magneto--optical experiments 
performed so far in 
magnetic metals, insulators, and semiconductors
observed magnetization dynamics
that could be  interpreted in terms of light--induced 
time--dependent thermal effects.
Following the observation of Ref.\onlinecite{bigot1},  
many works 
focussed on ultrafast light--induced demagnetization, which 
involves the 
time--dependent 
collective magnetization {\em amplitude}.  
The physical mechanisms that lead to 
quenching of the magnetization within a picosecond or less
in materials ranging from transition metals 
to III(Mn)V semiconductors are still under debate, 
but are mostly believed to be triggered by transient 
changes in the carrier and spin effective temperatures. 
\cite{bigot1,bigot2,bigot3,koop1,koop2,kojima-03,wang-demag,wang-rev}
However, transient magnetic effects 
have also been observed in the  initial non--thermal
temporal regime, where 
the concept of carrier temperature is not meaningful.
\cite{bigot2,bigot3}
Light--induced changes in the magnetization {\em orientation} 
have also been observed in both metals and semiconductors 
and mostly attributed to transient changes 
in the magnetic easy axes due to 
temperature elevation. 
\cite{bigot-chemphys,bigot-05,kampen,kimel-anis,munekata,munekata1,tolk}
Such thermal quasi--equilibrium effects  induce a 
magnetization precession 
with a period $\sim$100ps.

In most of the above experiments, the observed  magnetization changes 
result from the transient temperature rise following 
optical absorption and carrier relaxation. 
However, there is a limit in the magnetization speed that can 
arise from 
incoherent processes largely based on the heating of the magnetic system. 
Far more desirable is magnetization control 
based on {\em coherent} and non--thermal physical processes. 
The femtosecond coherent temporal regime  offers the most flexibility for fast 
magnetization control limited only by the optical 
pulse duration.
 Coherent manipulation of the magnetization was
demontrated experimentally in magnetic 
dielectrics and insulators \cite{kimel-coh-1,kimel-coh-2,kimel-rev}, 
while the interplay of coherent excitation and spin--orbit interaction 
in magnetic insulators  was addressed theoretically in 
Refs. \onlinecite{hub-1,hub-2}. 

Compared to other magnetically--ordered materials, 
III-Mn-V ferromagnetic semiconductors offer 
certain advantages and new possibilities for ultafast magnetization control and dynamics.
These advantages stem from the carrier--induced nature of 
the magnetic order and the clear distinction between localized (Mn) 
and itinerant (valence band hole) spins. 
Static measurements 
have shown that  III-Mn-V heterostructures 
 are highly sensitive to external stimuli  
such as  electrical gate,  currents, or light. 
\cite{ohno-2000,chiba,yaman,koshihara}
A light--induced  out-of-plane magnetization rotation 
towards the direction of propagation of a circularly polarized  
optical field perpendicular to the ground state magnetization 
was reported in Refs.  \onlinecite{oiwa-1,oiwa-2}
for Ga(Mn)As epilayers.
Later ultrafast experiments interpreted their findings in terms of  
photoexcited carrier spin 
\cite{kimel-04} 
and thermal \cite{kojima-03,wang-demag} 
effects (for a review see Ref. \onlinecite{wang-rev}). 

Recently, Wang {\em et.al.} \cite{wang-remag} reported an enhancement 
of the magnetization amplitude and the ferromagnetic order in GaMnAs
induced by  the photoexcited hole population. 
This enhancement occurs  on a $\sim$100ps time scale, 
following the initial subpicosecond demagnetization 
\cite{wang-demag}
and thermalization.
More recently, Wang {\em et.al.} \cite{wang-08} 
reported 
the first observation of two temporal 
regimes of magnetization dynamics in III-Mn-V semiconductors. 
The first regime lasts 
for a few hundreds of fs and is governed by a {\em quasi--instantaneous} 
tilt of the collective magnetization in  response to optical excitation
at high energies ($\sim$ 3.1eV).
In this femtosecond regime, a photoinduced 
four--state ferromagnetic hysterisis 
was measured, which
implies femtosecond detection of magnetic memory states. 
This initial magnetization dynamics is clearly distinguished 
from the subsequent thermal regime, which is 
governed by magnetization precession 
on the 100ps timescale around the magnetic easy axes.

The observations  of Ref.\onlinecite{wang-08} point out the need 
for a microscopic theory of collective spin dynamics in 
the initial coherent regime of III(Mn)V semiconductors, which 
treats the nonlinear response of the 
magnetization to the ultrafast optical excitation. 
The theoretical prediction of a light--induced 
 magnetic interaction resulting in a Kondo resonance 
in  the nonlinear optical response 
and the pump--probe spectra of doped semiconductors 
was reported in Ref.\onlinecite{kondo}. 
This light--induced many--body effect is generated by a 
second order Raman--like process and should be most pronounced in the 
case of below--resonance photoexcitation in the transparency regime, 
where heating effects are suppressed.   
Ref. \onlinecite{pier-04}
suggested the possibility of inducing 
ferromagnetic order by exciting 
undoped paramagnetic II-Mn-VI semiconductors
well below the optical absorption threshold. 
A brief description of a microscopic mechanism 
for coherent ultrafast  magnetization dynamics  in 
III-Mn-V ferromagnetic semiconductors 
was presented  in Ref. \onlinecite{chovan}, 
while Ref. \onlinecite{sham-07} addressed the 
subsequent incoherent regime and attributed the 
ultrafast demagnetization 
to the scattering 
of the Mn spins with 
spin--flip excitations of the hot hole Fermi sea. 

In this paper 
we develop 
in detail a theory 
that describes the  ultrafast nonlinear
response 
of the collective  spins
in ferromagnetic semiconductors.
We present calculations 
for the most basic Hamiltonian that applies to a wide 
range of  ferromagnetic semiconductor 
systems.
The magnetic exchange interaction
and the coupling to the optical field  
are treated 
within the mean field approximation, which 
has been shown to describe
well  the 
ground state and thermodynamic properties of 
III-Mn-V  semiconductors  \cite{mean,review}
as well as the ultarafast nonlinear optical 
response of semiconductors for strong photoexcitation. 
 \cite{sbe,sbe-sham}
In Section \ref{setup} we set up 
the problem at hand, 
while in Section \ref{sec:eom}
we derive the mean field equations of motion 
in the coherent limit 
in the case of several coupled valence bands. 
The dephasing effects due to the mixing of the hole spin states 
are described in Section \ref{sec:relax}
with the Lindblad 
semigroup  method. \cite{lindblad}
This method allows us to 
treat consistently both 
hole spin  relaxation/dephasing 
and interband polarization dephasing 
resulting from hole spin--flip interactions.  
In Section \ref{sec:single}
we extract a simple one--band model from the full theory 
and use it in Section \ref{sec:numer} to calculate the 
light--induced 
Mn  spin
trajectories and 
discuss the role of hole spin and interband polarization 
dephasing, magnetic anisotropy,  
and photoexcitation intensity and duration. 
Our numerical results predict 
an ultrafast  tilt of the Mn spin away from its initial (equilibrium) 
value, which develops  on a timescale comparable to the optical 
pulse duration.
The magnitude of this tilt 
is controlled by the optical pulse intensity and duration. 
The direction of the tilt and overall shape of the Mn spin trajectory 
depend sensitively on the hole spin dephasing  and on the  interband polarizations.
In Section \ref{sec:inter}  we interpret the above numerical results 
by deriving from the full theory an effective Landau--Gilbert--like \cite{landau} 
equation of motion for the Mn spin using the adiabatic 
following approximation. 
We show that the effective magnetic anisotropy  fields 
that govern the precession and relaxation 
within the above 
 Landau--Gilbert picture acquire a 
time--dependent correction 
determined by the optical pulse amplitude,
the photoexcited interband polarizations, and the mixing of the hole spin states.  
This  correction results in  magnetization  dynamics 
on a timescale comparable to the optical pulse duration
and may be interpeted in terms of light--induced 
precession and relaxation. 
After the pulse is gone, the Mn spin trajectory  is controlled by the 
magnetic anisotropies due to the 
Fermi sea (thermal) carriers. If the initial magnetization response to 
the optical excitation results 
in  a sufficiently small tilt, the Mn spin  precesses 
around its initial configuration, with a period determined by the 
zero--momentum magnon energy, as  described by 
linearizing the equations of motion.  
On the other hand, for sufficiently high photoexcitation intensity, 
the initial light--induced magnetization tilt 
brings the magnetic system sufficiently far from the 
magnetic energy minimum 
so that nonlinear magnetic effects become important. 
 In this case, the Mn spin evolution
differs from a simple precession.  
Our theory predicts two distinct temporal regimes of magnetization 
evolution, the first of which is governed by the optical excitation 
while the second is governed by the magnetic anisotropies
due to the thermal carriers. 
We discuss the signatures of the above  transient and nonlinear 
magnetic effects in ultrafast magneto--optical  pump--probe spectroscopy
and end with our conclusions in Section \ref{sec:concl}. 
The details of our calculations are presented 
in the five Appendices.

\section{Problem setup}
\label{setup}

We start with the Hamiltonian \cite{mean,review} 
$ H(t)= H +H_{L}(t)$, 
where $H= K_{e}+ K_h + H_{\rm{exch}}$ 
is the Hamiltonian in the absence of optical excitation and 
$H_L(t)$ describes the coupling of the optical fields.
\begin{equation} 
K_e=\sum_{{\bf k} n} \varepsilon^c_{{\bf k} n} 
\hat{e}^\dag_{{\bf k} n} 
\hat{e}_{{\bf k} n} \ , \ 
K_h=\sum_{{\bf k} n} \varepsilon^v_{-{\bf k} n} 
\hat{h}^\dag_{-{\bf k} n} 
\hat{h}_{-{\bf k} n}
\end{equation} 
describe the electron and hole band energies. 
The conduction band electrons are  
created by the operator 
$\hat{e}_{{\bf k}n}^{\dag}$,  
where ${\bf k}$ is the momentum and  
$n$ labels different conduction bands. 
 $\varepsilon^{c}_{{\bf k} n}$
is the $n$--band dispersion (we set $\hbar$=1).
The valence holes are 
created by the operator 
$\hat{h}_{{\bf k} n}^{\dag}$, where 
${\bf k}$ is the hole momentum and 
$n$ the band index.
Their 
dispersion 
$\varepsilon^{v}_{{\bf k} n}$ is determined by  the  bandstructure. 
The Mn impurities act as acceptors, which create 
a hole Fermi sea in the valence band, and also
provide  randomly distributed  $S=5/2$  
 spins, 
${\bf S}_j$, that are localized at positions ${\bf R}_j$. 
These local moments 
interact  with the  hole spin
via the antiferromagnetic Kondo--like  exchange interaction 
\begin{equation}
\label{Hexch}
H_{\rm{exch}} = \frac{\beta}{V}
\sum_{j {\bf k} {\bf k'} n n^{\prime}} 
{\bf S}_j \cdot {\bf s}_{n n^{\prime}}
e^{i ({\bf k} - {\bf k}^\prime) \cdot {\bf R}_j} 
\hat{h}^{\dag}_{-{\bf k}n} \hat{h}_{-{\bf k}^{\prime} n^{\prime}},
\end{equation}
where ${\bf s}$ 
is the hole spin operator,  $V$ is the volume, and 
$\beta$ is the exchange constant.
Finally, the coupling of the optical pulse
is described by 
\begin{equation}
\label{HL-full} H_{L}(t)= -
\sum_{n n^{\prime} {\bf k}} 
d_{n n^{\prime}}(t)  
\hat{e}^{\dag}_{{\bf k}n} \hat{h}^{\dag}_{{-\bf k}n^{\prime}}
+ h.c,
\end{equation}
where $d_{n n^{\prime}}(t)={\mu}_{n n^\prime} {\cal E}(t)$ 
is the Rabi energy,   ${\cal E}(t)= {\cal E}
\exp{[ -t^{2}/\tau_p^{2}]}$ 
is the optical  (pump) pulse, with duration 
$\tau_p$, and  $\mu_{ n n^{\prime}}$ is the dipole transition 
matrix element
between the valence band $n^{\prime}$ 
and the conduction band $n$.
The optical pulse propagates along the growth direction 
z,  which is 
perpendicular to the ground state spins.

The ground state, thermodynamic, and transport properties 
of III(Mn)V semiconductors in the metallic regime 
(hole densities $\sim$10$^{20}$cm$^{-3}$) 
are well described  by treating 
the magnetic exchange interaction, Eq.(\ref{Hexch}), 
within the 
mean field  virtual crystal approximation. 
\cite{review}
This approximation
neglects spatial correlations 
and assumes 
uniformly distributed classical Mn spins, 
justified in the limit $S\rightarrow\infty$, 
where $S$ is the Mn spin amplitude. 
The 
holes then experience an effective  magnetic field 
proportional to the 
Mn spin.
The valence band  splits into two spin--polarized 
bands separated 
by the magnetic exchange energy
$\Delta = \beta c S$, where $c$ is the density of the localized Mn spins.
The typical values of $\Delta$ observed in the 
GaMnAs and InMnAs ferromagnetic 
semiconductors that exhibit the highest 
critical temperatures
are comparable 
to the Fermi energy, $E_F \sim$100meV,  of the ground state
hole Fermi sea.

In the ground state, the 
Mn spin
points along the easy axis direction in order to minimize the 
total energy of the Fermi sea carriers:
\begin{equation} 
E_h({\bf S}) = \sum_{{\bf k} i} E^h_{{\bf k} i} 
n_{{\bf k} i},  
\end{equation} 
where 
$E^h_{{\bf k} i}({\bf S})$ are the eigensvalues of 
the Hamiltonian 
$K_h + H_{exch}$ for given Mn spin 
${\bf S}$ 
and 
$n_{{\bf k} i}$
are the populations of the corresponding eigenstates. 
In  III-Mn-V semiconductors, 
this mean field  total energy 
depends strongly on the orientation of 
${\bf S}$.\cite{macd-anis,dietl} This anisotropy is believed to mainly arise 
from  the valence bandstructure, in particular the 
spin--orbit coupling of different valence bands, which was taken into account 
in Refs.\onlinecite{macd-anis,dietl} 
within the ${\bf k} \cdot
  {\bf p}$ envelope function approximation.
The  easy axis 
depends sensitively on the hole  
distribution among different valence bands. 
The calculated $E_h({\bf S})$ is well described 
by an expansion in terms of  ${\bf S}$. \cite{macd-anis} 
The following expansion
describes the  static
experimental measurements:
\cite{welp}
\begin{equation} 
E_h
 = K_{c} (\hat{S}_x^2 
\hat{S}_y^2 + \hat{S}_x^2 
\hat{S}_z^2 + \hat{S}_y^2 
\hat{S}_z^2) 
+ K_u \hat{S}_x^2 + 
K_{uz} \hat{S}_z^2,
\label{tot-en}
\end{equation} 
where 
$\hat{{\bf S}}={\bf S}/S$, 
 $K_{c}$ 
is the lowest cubic anisotropy constant, and  
$K_u$ and $K_{uz}$ 
are the first order uniaxial anisotropy constants, 
attributed to strain, whose origin is
still being debated. \cite{hamaya}
For $K_u=K_{uz}=0$, the ground state magnetization 
is either parallel to the axes $\pm x, \pm y$ or $\pm z$, 
if $K_c >0$, or 
points along the diagonals  $|x|=|y|=|z|$,
if $K_c<0$. 
For the parameters of interest in Ga(Mn)As epilayers, 
$K_c<0$, and the easy axis direction is determined 
by the magnitude and sign of the uniaxial anisotropy constants 
$K_u$ and $K_{uz}$. 
A sufficiently large $K_{uz}>0$ ensures  that
the ground state magnetization lies within the $x$-$y$ plane
as observed experimentally. 
The observed temperature and hole concentration 
dependence of the easy axis  suggests that 
$K_u>0$. \cite{welp} 
For $K_u>|K_c|$, the easy axis  points along the 
$\pm y$ axis, while for  
$K_u<|K_c|$ it points at an angle $\phi$ from the $x$--axis,
where $\cos 2 \phi = K_u/|K_c|$.

Even though ${\bf k}\cdot{\bf p}$ and mean field 
theory explain the main anisotropy effects, the 
interpretation of the  observed changes of the 
in--plane easy axis with temperature and hole concentration 
require further theoretical investigations. 
A complete microscopic theory that includes the magnetic anisotropy 
becomes even more complicated in the case of 
ultrafast optical excitation. As observed experimentally, \cite{wang-08} 
the magnetization responds to the
photoexcitation of high energy ($\sim$3.1eV) valence bands states, which are  
empty in the ground state and lie far  from the Brillouin zone center,  
well before carrier or spin thermalization. 
In this initial highly nonequilibrium  regime, 
a fully microscopic theory must address both the valence bandstructure 
at $\sim$3eV energies and the  coherent and non--thermal 
effects. 
Here we address the general magnetic semiconductor system and 
 include the magnetic anisotropy effects by using the Fermi sea 
energy expression Eq.(\ref{tot-en}). The time dependent response of 
 the Mn spin 
${\bf S}$  to the nonlinear optical excitation, discussed 
in the next section,  
changes $E_h({\bf S})$, which  leads to an additional complex nonlinear 
magnetization dynamics.

\section{Equations of Motion }
\label{sec:eom}

To describe the ultrafast optical and spin response, we proceed
in two steps. In this section  we derive the mean field equations of motion, 
while in the next section 
we derive the dephasing contributions 
due to the 
hole spin--flip 
interactions.
The components of the density matrix
$\langle \rho \rangle $ 
 are obtained from the equations of motion 
\begin{equation} 
i \partial_t \langle \rho \rangle = 
\left. \langle [  \rho , H(t)] \rangle \right|_{HF} 
+ i \partial_t\left. \langle \rho \rangle \right|_{relax},
\label{eom} 
\end{equation} 
where the last term describes 
the dephasing and relaxation contributions.
By factorizing all higher density matrices (Hartree--Fock
 approximation),  which couple due to the 
many--body exchange interaction, 
we obtain  a closed system of equations
for the optical
polarizations, 
spins, and  carrier
populations and coherences. 

We are interested in calculating the  macroscopic 
magnetization measured in ultrafast pump--probe 
magneto--optical experiments. This is dominated 
by the contribution of 
the average Mn spin 
\begin{equation} 
{\bf S} = \frac{1}{c V} \sum_{i} \langle {\bf S}_i \rangle.
\end{equation} 
From Eq.(\ref{eom}) we obtain within the mean field approximation 
the equation of motion that describes the magnetization dynamics: 
\begin{equation} 
\partial_{t} {\bf S}
= {\bf S} \times \left[ {\bf H} -
\frac{\beta}{V} \sum_{{\bf k}}{\bf s}^h_{{\bf k}}\right], 
 \label{Mn-spin} 
\end{equation} 
where ${\bf H}$ is a magnetic field. 
The right hand side (rhs) of the above 
equation 
describes the precession of the Mn spin around 
an effective  time--dependent
magnetic field
determined by  ${\bf H}$ and the 
 mean hole spin
\begin{equation} 
{\bf s}_{{\bf k}}^h
= 
\sum_{nn^{\prime}} 
{\bf s}_{nn^{\prime}}
\langle 
\hat{h}^{\dag}_{-{\bf k}n} \hat{h}_{-{\bf k}
  n^{\prime}}\rangle, 
\label{h-spin} 
\end{equation}
where ${\bf s}_{nn^{\prime}}$ 
are the matrix elements of the hole spin operator
${\bf s}$ between the valence band eigenstates.  
Eq.(\ref{Mn-spin}) 
corresponds
to the Landau--Gilbert 
picture of magnetization dynamics \cite{landau} 
and conserves  
the amplitude of ${\bf S}$.

The hole spin,  populations, and  inter--valence band coherences  
are described
by the density matrices 
$\langle \hat{h}^{\dag}_{-{\bf k}n} 
\hat{h}_{-{\bf k} n^{\prime}}\rangle$, 
whose 
equations of motion read 
 \begin{eqnarray} 
&& i \partial_t 
\langle \hat{h}^{\dag}_{-{\bf k}n} 
\hat{h}_{-{\bf k} n^{\prime}}\rangle
=
\left(\varepsilon^v_{{\bf k} n^{\prime}} 
- \varepsilon^v_{{\bf k} n} \right) 
\langle \hat{h}^{\dag}_{-{\bf k}n} 
\hat{h}_{-{\bf k} n^{\prime}}\rangle 
\nonumber \\
&& + \beta c
\sum_{m^{\prime}}{\bf S} \cdot 
\left[ 
{\bf s}_{n^{\prime} m^{\prime}} 
\langle \hat{h}^{\dag}_{-{\bf k}n} 
\hat{h}_{-{\bf k} m^{\prime}}\rangle -
 {\bf s}^*_{n m^{\prime}} 
\langle \hat{h}^{\dag}_{-{\bf k} m^{\prime}} 
\hat{h}_{-{\bf k} n^{\prime}}\rangle \right]
\nonumber \\
&& + \sum_{m^{\prime}}
d^*_{m^{\prime}n}(t) 
\langle \hat{h}_{-{\bf k}
n^{\prime}} \hat{e}_{{\bf k} m^{\prime}} 
\rangle
- \sum_{m^{\prime}}
d_{m^{\prime} n^{\prime}}(t) 
\langle  \hat{h}_{-{\bf k}
n} \hat{e}_{{\bf k} m^{\prime}}
 \rangle^*
\nonumber \\
&& + i \left. \partial_t \langle \hat{h}^{\dag}_{-{\bf k} n} 
\hat{h}_{-{\bf k} n^{\prime}}\rangle \right|_{relax}.
 \label{dm-h} 
\end{eqnarray}
The second line on the rhs of the above equation describes 
the change in the hole
states
due to the exchange interaction. 
The third line describes the excitation 
of hole coherences and populations by the optical pulse, via second--order 
Raman processes. 
Similarly, we obtain for the electron
populations and coherences  
 \begin{eqnarray} 
&& i \partial_t 
\langle \hat{e}^{\dag}_{{\bf k}n} 
\hat{e}_{{\bf k} n^{\prime}}\rangle
=
\left(\varepsilon^c_{{\bf k} n^{\prime}} 
- \varepsilon^c_{{\bf k} n} \right) 
\langle \hat{e}^{\dag}_{{\bf k}n} 
\hat{e}_{{\bf k} n^{\prime}}\rangle 
\nonumber \\
&& 
 + \sum_{m^{\prime}}
d^*_{n m^{\prime}}(t) 
\langle \hat{h}_{-{\bf k}
m^{\prime}} \hat{e}_{{\bf k} n^{\prime}} 
\rangle
- \sum_{m^{\prime}}
d_{n^{\prime}m^{\prime}}(t) 
\langle  \hat{h}_{-{\bf k}
m^{\prime}} \hat{e}_{{\bf k}n }
 \rangle^* \nonumber \\
&&+ i \left. \partial_t \langle \hat{e}^{\dag}_{{\bf k} n}
\hat{e}_{{\bf k} n^{\prime}}\rangle \right|_{relax}.
 \label{dm-e} 
\end{eqnarray}
The total carrier populations
with given momentum ${\bf k}$, 
\begin{equation} 
N^e_{{\bf k} } 
= \sum_{n}  \langle \hat{e}^{\dag}_{{\bf k}n} 
\hat{e}_{{\bf k} n}\rangle
\ , \ 
N^h_{{\bf k}} 
= \sum_{n}  \langle \hat{h}^{\dag}_{-{\bf k}n} 
\hat{h}_{-{\bf k} n}\rangle,
\end{equation} 
 are not affected by the magnetic exchange 
interaction, 
which  within the mean field 
approximation only changes the spin.
They satisfy 
equations of motion similar to the  Bloch equations: 
 \begin{equation} 
\partial_t 
N^{h}_{-{\bf k}} 
=
2 Im \sum_{n n^{\prime}} 
d^*_{n^{\prime}n}(t) 
\langle \hat{h}_{-{\bf k}
n} \hat{e}_{{\bf k} n^{\prime}} 
\rangle
- \left( N^h_{-{\bf k}} 
- f^{h}_{-{\bf k}} \right)/T_1^h
 \label{N-h} 
\end{equation}
and 
 \begin{equation} 
\partial_t 
N^{e}_{{\bf k}} 
=
2 Im \sum_{n n^{\prime}}
d^*_{n^{\prime}n}(t) 
\langle \hat{h}_{-{\bf k}n} \hat{e}_{{\bf k} n^{\prime}} 
\rangle
-  N^e_{{\bf k}}/T_1^e, 
 \label{N-e} 
\end{equation}
where $T^h_1$ and $T^e_1$ 
are the spin--independent population relaxation times, 
due to the carrier--carrier and 
carrier--phonon
scattering processes which thermalize the 
carrier system.
The hole population  relaxes  to the  thermal 
 distribution 
$f^{h}_{{\bf k}}$.

Using Eq.(\ref{h-spin}) 
for the total hole spin,
we obtain from 
Eq.(\ref{dm-h})  
after some algebra that
 \begin{equation} 
 \partial_t {\bf s}^h_{{\bf k}}=
\beta c{\bf S}\times{\bf s}^h_{{\bf k}}
+Im {\bf h}_{{\bf k}}(t) 
+ i
\langle [K_h,{\bf s}^h_{{\bf k}}] \rangle
+ \left. \partial_t {\bf s}^h_{{\bf k}} \right|_{relax}.
\label{hole-spin} 
\end{equation} 
The first term on the rhs  describes the 
precession of the hole spin around the mean field 
created by the Mn spin.
 The second term 
describes the 
photoexcitation of hole spin via second--order 
Raman processes, which are 
determined by the 
transition matrix elements (selection rules)  
and the interband  optical polarizations:
\begin{eqnarray} 
{\bf h}_{{\bf k}}(t) = 
 2  \sum_{n n^{\prime} m^{\prime}}
 d_{n m^{\prime}}^*(t) {\bf s}_{m^{\prime} n^{\prime}}
\langle \hat{h}_{-{\bf k}
n^{\prime}} \hat{e}_{{\bf k} n} 
\rangle.
\label{h-def}
\end{eqnarray}
The third term on the rhs of Eq.(\ref{hole-spin}) 
arises when the valence band 
states are not eigenstates of the hole spin:
$[{\bf s}^h_{{\bf k}},K_h] \ne 0$.
In this case, 
the 
hole spin dynamics depends 
on the valence band energies and
is determined by the individual 
coherences, Eq.(\ref{dm-h}),  
between the 
different valence bands. 
In III-Mn-V semiconductors, 
the hole spin is not conserved  due to the strong spin--orbit interaction.
Recalling Eq.(\ref{Mn-spin}), 
we  conclude that the mixing of the hole spin states
and the valence bandstructure modifies the effective magnetic field experienced by the Mn spin
as described by the 
third term on the rhs of Eq.(\ref{hole-spin}).

The photoexcitation of carrier spin, which
triggers the  magnetization dynamics, 
is governed
by the interband optical polarizations 
\begin{equation} 
P_{{\bf k} nn^\prime} 
= \langle \hat{h}_{-{\bf k} n} 
\hat{e}_{{\bf k} n^{\prime}} \rangle, 
\label{pol-def} 
\end{equation}
determined by the equations of motion 
\begin{eqnarray} 
&&i \partial_{t} P_{{\bf k} nn^\prime}
= \left(\varepsilon^c_{{\bf k} n^\prime} 
+\varepsilon^v_{{\bf k} n} - \omega_p 
-i/T_2 \right) P_{{\bf k} nn^\prime}
\nonumber \\
&&
+ \beta c {\bf S} \cdot
\sum_{m} {\bf s}_{nm} P_{{\bf k} m n^\prime} 
\nonumber \\ && 
- d_{n^\prime n}(t) \left[1 - \langle \hat{h}^\dag_{-{\bf k}
 n} 
\hat{h}_{-{\bf k} n} \rangle 
- \langle \hat{e}^\dag_{{\bf k} n^\prime}
\hat{e}_{{\bf k} n^\prime} \rangle \right] \nonumber \\ 
&&
+ \sum_{m \ne n}
 d_{n^\prime m} 
\langle \hat{h}^\dag_{-{\bf k} m} \hat{h}_{-{\bf k} n} \rangle 
+  \sum_{m\ne n^\prime}d_{m n} 
\langle \hat{e}^\dag_{{\bf k} m} \hat{e}_{{\bf k} n^\prime}
\rangle 
\nonumber \\
&& + i \partial_t  
\left. P_{{\bf k} n n^\prime}\right|_{relax}.
\label{P}
\end{eqnarray} 
In the above equation, 
 $\omega_p$ is the pump optical pulse central frequency   and 
$T_2$ describes the spin--independent 
contribution to the  polarization dephasing. $T_2$ can be 
quite short  
due to the disorder, which relaxes the momentum conservation by introducing a 
one--body  potential. For weak disorder, this effect can be treated by introducing a
dephasing time comparable to the 
momentum scattering time. 
The second line on the rhs of the above equation 
is due to the change of the hole states induced by  the effective
 magnetic field $\beta c {\bf S}(t)$. 
The third line 
describes the  Pauli--blocking nonlinearity
(Phase Space Filling) \cite{sbe},
while 
the fourth line 
describes the contribution of carrier coherences between
 the different bands. Finally, the last line describes the 
spin--dependent polarization dephasing due to the hole spin--flip
 interactions, discussed in the next section.
We note from the above equations 
that the interband polarizations, valence band  coherences, and hole states  
depend on  the effective magnetic
field proportional to  ${\bf S}(t)$. 
This spin can deviate significantly  from its ground state configuration 
in the case of strong photoexcitation, which in turn 
changes the hole states
as compared to the ground state. 
Such light--induced
deviations from equilibrium  increase with photoexcitation intensity
 and are  
described non--perturbatively  by solving numerically the above 
system of 
coupled equations
without expanding in terms of the optical field.

\section{Hole Spin and 
Polarization Dephasing} 
\label{sec:relax}

\subsection{Lindblad formalism} 

The 
equations derived in the previous section do not include
dephasing and relaxation contributions.
Within the  semiconductor Bloch Equations, \cite{sbe} 
such effects are treated phenomenologically 
by introducing 
effective dephasing and relaxation times
to the polarization and population equations of motion. 
Here 
we must  also consider 
the dynamics of the carrier and Mn spins, 
which within the mean field approximation 
precess around each other while conserving their magnitudes. 
In the III-Mn-V system,
the hole 
 spin relaxation is strong, mainly due to the spin--orbit interaction.
For example, 
the combination of 
spin--orbit coupling
in the valence band
and  disorder--induced scattering between momentum states,  
leads to  hole spin dephasing and relaxation times 
of the order of  10's of fs.
 \cite{jungwirth,zutic}
Such times are  comparable 
to the hole spin precession period around the Mn spin  
and cannot be neglected.

Carrier spin relaxation is often described within 
the spin Bloch equations. 
 \cite{zutic,halperin}
However, hole spin relaxation
also leads to interband 
polarization  dephasing, 
and both effects must be treated on equal footing  for our purposes here.  
For this 
we use the Lindblad 
semigroup description
of dissipative quantum dynamics. \cite{lindblad}
Under the  general assumptions of linear coupling between bath and
system operators and 
Markovian/relaxation time approximation,  
as well as density matrix  positivity 
and semigroup--type time evolution, 
the relaxation contribution to the density matrix equation of motion 
can be expressed in the form 
 \cite{lindblad} 
\begin{eqnarray}
\label{lind}
\left. \partial_t \rho \right|_{relax} &= & 
\Gamma_{\perp} \sum_{{\bf k}^\prime m^\prime } 
\left[  2 \langle 
 L^\dag_{{\bf k}^\prime m^\prime }  \rho L_{{\bf k}^\prime m^\prime } 
\rangle \right. \nonumber \\
&& \left.- \langle L_{{\bf k}^\prime m^\prime }^{\dag}
 L_{{\bf k}^\prime m^\prime } \rho \rangle - \langle \rho 
L_{{\bf k}^\prime m^\prime }^{\dag} L_{{\bf k}^\prime m^\prime } 
\rangle \right] , 
\end{eqnarray}
where 
$\Gamma_{\perp}$ is the spin dephasing rate. 
The Lindblad operators 
$L_{{\bf k} m}$
must be chosen to describe the relaxation processes at hand. 
In our case, the hole spin must relax towards the direction
antiparallel to the Mn-spin  ${\bf S}(t)$.  
As we show below, this can be achieved by choosing 
\begin{equation} 
L_{\bf{k} m} = \hat{h}^{\dag}_{-{\bf k} m \Downarrow} 
\hat{h}_{-{\bf k} m \Uparrow},
\label{lind-op}
\end{equation} 
where  $\hat{h}^{\dag}_{\bf{k} m \Uparrow}$ 
( $\hat{h}^{\dag}_{\bf{k} m \Downarrow}$) 
creates a hole with 
 spin parallel (anti--parallel)
to  the Mn spin ${\bf S}(t)$ and the  index $m$ labels the different 
basis states.

\subsection{Hole spin dephasing and relaxation} 

First we derive the hole spin dephasing and relaxation 
described by the
Lindblad  operator Eq.(\ref{lind-op}). 
Noting that the hole spin relaxes to 
a direction antiparallel to the Mn spin, 
it is useful to introduce its components 
parallel and perpendicular 
 to the  unit vector ${\bf \hat{S}}$: 
\begin{equation} 
{\bf s}^{h}_{{\bf k}m} 
= {\bf s}^{h}_{{\bf k}m{\perp}}
+ \hat{{\bf S}}  s^{h}_{{\bf k}m \parallel},
\label{decomp} 
\end{equation} 
where $
 s^{h}_{{\bf k} m \parallel}
=  {\bf \hat{S}} \cdot {\bf s}^{h}_{{\bf k} m}
$
and 
\begin{equation} 
{\bf s}^{h}_{{\bf k}m{\perp}}
=  {\bf \hat{S}} \times \left( 
{\bf s}^{h}_{{\bf k}m} \times {\bf \hat{S}} \right).
\label{perp} 
\end{equation} 
As derived in Appendix \ref{spin-deph}, 
the  Lindblad operator Eq.(\ref{lind-op}) gives the 
following spin--Bloch equations: 
\begin{eqnarray}
\partial_t \left. {\bf s}^{h}_{{\bf k}m{\perp}}
\right|_{relax}= - \Gamma_{\perp}
{\bf s}^{h}_{{\bf k}m{\perp}},
\label{spineom-perp} \\
\partial_t \left. {\bf s}^{h}_{{\bf k} m \parallel}
\right|_{relax} 
=-\Gamma_{\parallel}\left(
{\bf s}^{h}_{{\bf k} m \parallel}
+ m^h_{{\bf k} m} \right),
\label{spineom-par}
\end{eqnarray}
where $\Gamma_{\parallel}= 2 \Gamma_{\perp}$ 
is the spin relaxation rate and  
\begin{eqnarray} 
m^h_{{\bf k} m} = 
N^{h}_{{\bf k} m}/2-(N^{h}_{{\bf k} m}/2)^2+({\bf s}^{h}_{{\bf k} m})^{2}
\end{eqnarray} 
is the quasi--equilibrium hole spin value.
$m^h$ corresponds to 
the maximum  spin, $s_{\rm{max}}$,  of  $N^h$ holes,  
given by the relation 
$s_{\rm{max}}
- s_{\rm{max}}^2= 
N^{h}/2-(N^{h}/2)^2$. 

\subsection{Spin--dependent Polarization Dephasing}

The hole spin dephasing   also  dephases 
the interband optical polarizations. 
To describe this, 
it is useful to use the basis of hole spin eigenstates 
discussed above  and define the
 interband electron--hole amplitudes 
\begin{equation} 
P_{{\bf k}m  n \sigma }
= \langle \hat{h}_{-{\bf k}m \sigma} 
\hat{e}_{{\bf k} n} \rangle,
\label{pol-spin-def} 
\end{equation}
where $\sigma=\uparrow, \downarrow$.
Using Eqs.(\ref{lind}) and 
(\ref{lind-op}),  
we derive
in Appendix (\ref{pol-deph})  
the following expression for the spin--dependent 
polarization dephasing:
\begin{eqnarray} 
&&\left. \partial_t P_{{\bf k} m n\uparrow} \right|_{relax} 
 = - \Gamma_{\perp} \times \nonumber \\
&&\left\{ P_{{\bf k}m n \uparrow}
\Bigg \lbrack
\frac{1}{2}  +   s^h_{{\bf k} m \parallel} 
+  \frac{\hat{{\bf S}}_z}{2}
( 1 - N^h_{{\bf k} m})
+  i ({\bf \hat{S}} \times {\bf s}^h_{{\bf k} m})_{z}
\Bigg \rbrack \right.
\nonumber \\
&& + \left. P_{{\bf k}m n \downarrow}
\Bigg \lbrack \frac{ \hat{{\bf S}}_{-}}{2}
( 1 -  N^h_{{\bf k} m}) 
+ i  ( {\bf \hat{S}} \times {\bf s}^h_{{\bf k} m} )_{-}
\Bigg \rbrack \right\}, \label{pol-deph-up} \\
&&\left. \partial_t P_{{\bf k} m n \downarrow} \right|_{relax} 
 = - \Gamma_{\perp} \times \nonumber \\
&&\left\{ P_{{\bf k} m n \downarrow}
\Bigg \lbrack
\frac{1}{2} + s^h_{{\bf k} m \parallel}  - \frac{ \hat{{\bf S}}_z}{2}
( 1 - N^h_{{\bf k} m} ) 
- i \left({\bf \hat{S}} \times {\bf s}^h_{{\bf k} m} \right)_{z}
\Bigg \rbrack \right.
\nonumber \\
&&+ \left. P_{{\bf k} m n \uparrow}
\left[ \frac{\hat{{\bf S}}_{+}}{2}
( 1 -  N^h_{{\bf k} m}) 
+ i \left( {\bf \hat{S}} \times {\bf s}^h_{{\bf k} m} \right)_{+}
\right] \right\}, 
\label{pol-deph-down} 
\end{eqnarray} 
where, for any vector ${\bf A}$, $ A_{\pm}= A_x \pm i A_y$. 
The above dephasing contribution 
depends nonlinearly on the photoexcitation,
via the
photoexcited hole contribution to the 
population $N^h$ and  spin 
${\bf s}^h$ and the light--induced changes in the Mn spin ${\bf S}$. 
These 
nonlinearities correspond to excitation--induced dephasing induced by 
hole spin--flip correlations.

\section{Single Band Approximation} 
\label{sec:single}

The general theory derived in the previous sections may be used to treat 
the bandstructure
relevant to the particular ferromagnetic material  of interest.  
Below  we discuss the general features 
of the dynamics, their physical origin,  and their sensitivity to 
different parameters by extracting 
from the general theory  
a simplified one--band model 
that captures the essential physics common in all materials.  
We therefore neglect 
bandstructure particularities, 
 such as  the nature of the 
high energy states far from the Brillouin zone center,
excited by the $\sim$3.1eV pump of Ref.\onlinecite{wang-08},  
or impurity bands.   
We assume that the magnetization dynamics 
is mainly triggered
by optical transitions between a single valence and conduction band of 
spin--$\uparrow$ and spin--$\downarrow$ states, 
whose mixing 
is described phenomenologically  with the Linbdlad 
approach. 
In the basis of carrier spin eigenstates  
and assuming a single band, 
Eq.(\ref{HL-full}) 
reduces to 
\begin{equation}
\label{HL} H_{L}= -
d_+  
\sum_{{ \bf k}}
\hat{e}^{\dag}_{{\bf k}\downarrow} \hat{h}^{\dag}_{{-\bf k}\uparrow}
- 
d_-  
\sum_{{ \bf k}}
\hat{e}^{\dag}_{{\bf k}\uparrow} \hat{h}^{\dag}_{{-\bf k}\downarrow}
+ h.c.\end{equation}
where $d_{\pm}(t) = \mu_{\pm} {\cal E}(t)$. 
The Rabi energy $d_+$ describes the coupling of the 
 right--circularly polarized component 
of the optical field, while $d_{-}$ describes the coupling of the 
left--circularly polarized component. The corresponding 
interband transition matrix elements 
$\mu_{\pm}$ depend on the bandstructure and  the admixture 
of spin--$\uparrow$ and spin--$\downarrow$ 
in the band states that mostly contribute to Eq.(\ref{HL-full}) 
for the energies of interest.  
Denoting 
\begin{equation} 
P^{+}_{{\bf k} \sigma} 
= \langle \hat{h}_{-{\bf k} \sigma}
\hat{e}_{{\bf k} \downarrow}  \rangle,
P^{-}_{{\bf k} \sigma} 
= \langle \hat{h}_{-{\bf k} \sigma}
\hat{e}_{{\bf k} \uparrow}  \rangle.
\end{equation}
and using Eqs.(\ref{pol-deph-up})  and (\ref{pol-deph-down}) and 
the results of Section \ref{sec:eom}, 
we obtain the following coupled equations of motion for 
the interband optical polarizations:
\begin{eqnarray}
&& i \partial_{t}  P^+_{{\bf k} \uparrow}- 
(\Omega_{{\bf k}} +  {\bf \Delta}_{{\bf k}z})  P^+_{{\bf k} \uparrow}
 -  {\bf \Delta}_{{\bf k}-}
P^+_{{\bf k} \downarrow} \nonumber \\
 &&= -d_{+}(t) \ (1 - n^e_{{\bf k}\downarrow} 
-  n^h_{{\bf k}\uparrow}),
\label{Pup1} \\
 &&i \partial_{t}  P^+_{{\bf k} \downarrow} 
-(\Omega_{{\bf k}} - {\bf \Delta}_{{\bf k}z}) 
 P^+_{{\bf k} \downarrow} -  {\bf \Delta}_{{\bf k} +} 
P^+_{{\bf k} \uparrow} \nonumber \\
&&=  d_{+}(t) \ {\bf s}_{ {\bf k} +}^h
+ d_{-}(t) \ {\bf s}_{ {\bf k} +}^e,
\label{Pdown1} \\
&& i \partial_{t}  P^-_{{\bf k} \downarrow}  
-(\Omega_{{\bf k}} -  {\bf \Delta}_{{\bf k}z}) 
 P^-_{{\bf k} \downarrow} -  {\bf \Delta}_{{\bf k} +} 
P^-_{{\bf k} \uparrow} \nonumber \\
&&= - d_{-}(t) \ ( 1 - n^e_{{\bf k} \uparrow} 
-  n^h_{{\bf k} \downarrow}),
\label{Pdown2}
\\ 
&&i \partial_{t}  P^-_{{\bf k} \uparrow}  - 
(\Omega_{{\bf k}} +  {\bf \Delta}_{{\bf k}z}) 
 P^-_{{\bf k} \uparrow} 
- {\bf \Delta}_{{\bf k} -}
P^-_{{\bf k} \downarrow} \nonumber \\
&&=  d_{-}(t) \ {\bf s}_{ {\bf k} -}^h +
d_{+}(t) \ {\bf s}_{ {\bf k} -}^e,
\label{Pup2} 
\end{eqnarray} 
where 
\begin{eqnarray}
\Omega_{{\bf k}} = 
\varepsilon^{v}_{\bf{k}} 
+ \varepsilon^{c}_{\bf{k}}  - \omega_p 
 - i \left[\frac{1}{T_2}
+ \Gamma_{\perp} \left( \frac{1}{2} + 
  s^h_{{\bf k} \parallel} \right)\right]
\label{Omega} 
\end{eqnarray}
gives the Mn spin--independent and 
\begin{eqnarray} 
{\bf \Delta}_{{\bf k}} = 
\frac{{\bf \hat{S}}}{2}  
\left[ \beta c S
-i \Gamma_{\perp}
( 1 - N^h_{{\bf k}}) \right] 
+  \Gamma_{\perp} {\bf \hat{S}} \times {\bf s}^h_{{\bf k}} 
\label{Delta} 
\end{eqnarray} 
the Mn spin--dependent 
contribution to the e--h pair energy and dephasing.
${\bf \Delta}_{{\bf k}}$
also determines the 
coupling between the  interband polarizations,
due to the mixing of the hole spins 
by the magnetic exchange interaction and 
the dephasing. 
The above e--h pair energies 
depend on
the light--induced 
deviations of 
the Mn and hole spins from their ground state configurations
and on the photoexcited hole populations, 
which give 
 nonlinear contributions to the optical polarization.
The rhs of Eqs.(\ref{Pup1}) and 
(\ref{Pdown2}) 
describes 
the  Pauli--blocking nonlinearities
(Phase Space Filling) \cite{sbe}, 
while the rhs of 
Eqs.(\ref{Pdown1}) and 
(\ref{Pup2}) 
describes the contribution 
of carrier spin coherences, ground state or photoexcited. 
The above polarization equations of motion treat 
the effects of the 
mixing of the spin--$\uparrow$ and spin--$\downarrow$ states 
by spin--orbit or other spin--flip interactions 
by using Eqs.(\ref{pol-deph-up}) and (\ref{pol-deph-down}).

Finally we turn to the equations of motion for the 
carrier spins. 
The hole spin dynamics is described by Eq.(\ref{hole-spin}).  
Within the one--band approximation, 
the mixing of the spin--$\uparrow$ 
and spin--$\downarrow$ hole states  
leads to the spin dephasing 
derived in  Appendix \ref{spin-deph}.
The second term on the rhs of Eq.(\ref{hole-spin})
describes the photoexcitation of hole spin
as 
determined by the 
interband polarizations and 
${\bf h}_{{\bf k}}(t)$, 
Eq.(\ref{h-def}).  
In the  basis of hole spin eigestates 
considered in this section, 
\begin{eqnarray} 
 {\bf h}_{{\bf k} x}(t)  
&=&   d^*_{+}(t) P_{{\bf k} \downarrow}^{+}(t)
+d_{-}^*(t) P_{{\bf k} \uparrow}^{-}(t)
\nonumber \\
{\bf h}_{{\bf k} y}(t) &=&  
-i \left[d^*_{+}(t)  P_{{\bf k} \downarrow}^{+}(t) 
-d_{-}^*(t)  P_{{\bf k} \uparrow}^-(t) 
\right], \nonumber \\  
{\bf h}_{{\bf k} z}(t) &=&  d^*_{+}(t) P_{{\bf k} \uparrow}^{+}(t) 
- d_{-}^*(t) P_{{\bf k} \downarrow}^{-}(t). \label{h} 
\end{eqnarray} 
The above equation describes 
a second--order nonlinear 
optical process
where the excitation of a spin--$\downarrow$ hole--spin--$\downarrow$ 
electron pair is followed by  the 
de--excitation of a  spin--$\uparrow$ hole--spin--$\downarrow$ electron 
 pair. 
This second--order (Raman) 
process is induced by the right--circularly polarized component of the 
optical field, 
while the left--circularly polarized component 
induces an analogous process involving 
the electron spin--$\uparrow$ states. 
 The above  process requires
a nonzero polarization $P_{{\bf k} \downarrow}^{+}$.  
This is possible on the one hand due to the coupling of the 
spin--$\uparrow$ and spin--$\downarrow$ 
hole states (and hence
$P_{{\bf k} \downarrow}^{+}$ with 
$P_{{\bf k} \uparrow}^{+}$, third term on the lhs of 
Eq.(\ref{Pdown1}))
described by ${\bf \Delta}_{{\bf k}}$, 
 due to 
the magnetic exchange interaction and the 
hole spin dephasing processes (e.g. spin--orbit), 
and on the other hand due to the presence of 
spin--$\uparrow$--spin--$\downarrow$ 
hole spin coherence 
in the ground state.
Finally, Eq.(\ref{h}) 
describes the photoexcitation of spin--polarized hole populations 
and a hole spin z--component 
via the 
excitation and subsequent de--excitation 
of a spin--$\uparrow$ hole 
and spin--$\downarrow$  electron 
 pair 
(right--circularly polarized component) 
or   a spin--$\downarrow$ hole 
and spin--$\uparrow$  electron 
pair 
(left--circularly polarized component).
The above {\em coherent} nonlinear  effects 
occur during the optical pulse and do not require the absorption of light.
They can also 
be induced by photoexciting the system below resonance, 
in the transparency
regime, provided that the photoexcitation intensity is sufficiently high 
to achieve an observable effect. Such below--resonance 
photoexcitation is advantageous since  undesirable 
effects such as  transient heating 
can be suppressed and thus the speed of a possible 
device can be maximized. 
The equation of motion for the electron spin has a 
form similar  to Eq.(\ref{hole-spin}): 
 \begin{eqnarray} 
\left( \partial_t + \Gamma^s_e \right)  {\bf s}^e_{{\bf k}}=
\ Im \ {\bf h}^e_{{\bf k}}(t) 
 \label{spin-e} 
\end{eqnarray}
where  $\Gamma^s_e$ is the electron spin dephasing rate and 
${\bf h}^e_{{\bf k}}$ 
describes the photoexcitation 
of conduction electron spin:  
\begin{eqnarray} 
 {\bf h}^e_{{\bf k} x}(t)  
&=&  d_{-}^*(t) P_{{\bf k} \downarrow}^{+}(t)
+ d^*_{+}(t) P_{{\bf k} \uparrow}^{-}(t),  \nonumber\\
 {\bf h}^e_{{\bf k} y}(t) &=&  
i \left[ d^*_{+}(t)  P_{{\bf k} \uparrow}^{-}(t) 
- d_{-}^*(t)  P_{{\bf k} \downarrow}^+(t) \right], \nonumber\\  
{\bf h}^e_{{\bf k} z}(t) &=& 
- [d^*_{+}(t) P_{{\bf k} \uparrow}^{+}(t) 
- d_{-}^*(t) P_{{\bf k} \downarrow}^{-}(t)].
 \label{e} 
\end{eqnarray}
In the next section we calculate the Mn spin dynamics
by solving the  above system of coupled equations
non--perturbatively, which allows us to 
treat large deviations of the Mn spin from 
its ground state configuration induced by strong 
photoexcitation.

\begin{figure}
\centerline{
\hbox{\psfig{figure=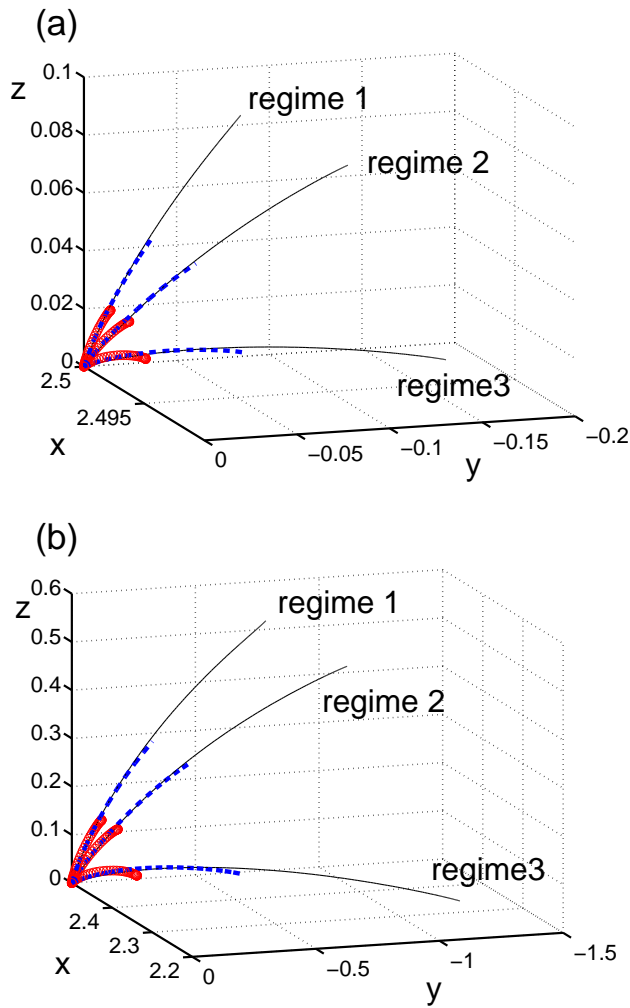,width=8.3cm}}
}
\caption{ (Color online) 
Mn spin trajectories without magnetic anisotropy 
in the three regimes of spin 
and polarization dephasing discussed in the text
for three pulse durations $\tau_p$  
and two different Rabi energies: 
(a) $d_{+}$(0)=20meV, (b) $d_{+}$(0)=60meV.
Thick circled line: $\tau_p$=100fs, Dashed line: $\tau_p$=250fs, 
Thin solid line: $\tau_p$=500fs.  
\label{Fig1}}
\end{figure} 

\section{Numerical Results} 
\label{sec:numer}

Within the mean field approximation, light--induced  Mn spin dynamics 
is triggered initially  by the photoexcitation of a 
hole spin component perpendicular to the 
ground state Mn spin. In the simple one--band 
approximation, 
the magnitude of such a  
spin component is determined by the ratio $d_{+}/d_{-}$ 
between the Rabi energies that describe  the coupling 
of the right-- and left--circularly 
polarized components of the optical field. 
This ratio is determined by the selection rules and the 
nature of the  bands that contribute to the 
magnetization dynamics. 
Within our simplified one--band model, 
$d_{+}/d_{-}$ also reflects the magnetic anisotropy of the system 
due to the bandstructure. In the absence of a 
complete theoretical understanding of such 
bandstructure effects,
especially for the high energy ($\sim$3.1eV) 
transitions observed experimentally to trigger ultrafast magnetization dynamics
in  Ga(Mn)As,\cite{wang-08} 
we consider here the extreme case of $d_{-}=0$, which 
corresponds to  right--circularly polarized light. 
For all the calculations presented in this paper, we consider a magnetic 
exchange energy $\beta c S$=125 meV  comparable to the 
Fermi energy $E_F$=100 meV,
fraction of  initial holes 0.33 of the Mn impurities, 
a hole mass $m_h= 7.15 m_e$, 
and a carrier thermalization 
time  $T_1$=165fs.
Our results are not very sensitive to the above 
parameters, with the exception of the exchange interaction $\beta$
that changes  the magnitude  of the effect. 
To make the connection between the calculated quantities and the experiment, 
we note that ultrafast magneto--optical pump--probe 
spectroscopy can be used to deduce the time evolution of the  z--component 
of the magnetization, $\sim S_z(t)$. \cite{wang-rev,wang-08}

We start with our results in the absence of magnetic anisotropy. 
 Fig.\ref{Fig1} shows the light--induced time evolution  of the Mn spin 
and its dependence on the hole spin and polarization dephasing, 
the Rabi energy, and the optical pulse duration. 
By comparing three different dephasing regimes
in Fig.\ref{Fig1}, 
one can see that the shape of the Mn spin trajectory depends sensitively
on the hole spin and polarization dephasing. 
Regime 1 corresponds to very short 
dephasing times, 
$1/\Gamma_{\perp}$=21fs and $T_2$=10.5fs, which are 
typical in III(Mn)V semiconductors. \cite{jungwirth}
Regime 2 corresponds to  $T_2$=330fs
with very short hole spin dephasing $1/\Gamma_{\perp}$=21fs. Finally,  
Regime 3 
neglects the spin dephasing altogether,  
$\Gamma_{\perp}=0$, $T_2$=330fs, which assumes that
the hole spin is a good quantum number.
As can be seen in Fig.\ref{Fig1}, the dephasing changes 
qualitatively the shape of the Mn spin trajectory, 
from the precession within the x--y plane expected in the absence 
of magnetic anisotropy to a complex magnetization tilt out of the x--y 
plane, in the direction of pulse propagation. 
The magnitude of this out--of--plane  tilt 
and the Mn spin z--component are 
enhanced by increasing $\Gamma_{\perp}$
or the pulse duration $\tau_p$ and by the short  $T_2$.
Furthermore, 
the  Mn spin component within the x--y plane 
rotates in a {\em clockwise} direction, 
opposite to the 
{\em counter--clockwise} direction expected 
for an effective magnetic field 
along the +z axis of pulse propagation. 
For right--circular--polarization, the latter
is the expected direction of the 
photoexcited hole spin in the absence of magnetic exchange interaction
or spin dephasing. 
However, our numerical results show that 
the photoexcited 
hole spin z--component 
is in fact {\em negative}, due to the    magnetic exchange interaction
and the mixing of the hole spin states. 
The dependence of the 
trajectories in Fig.\ref{Fig1} on the pulse duration 
shows that the light--induced Mn 
spin dynamics terminates soon after the end the 
photoexcitation, as expected in the absence of magnetic anisotropy.
Furthermore, the magnitude of the Mn spin tilt 
increases with photoexcitation intensity.  
The spin dynamics of Fig.\ref{Fig1} describes the 
fundamental 
response of the 
spin system to the optical excitation within the mean field approximation.

\begin{figure}
\centerline{
\hbox{\psfig{figure=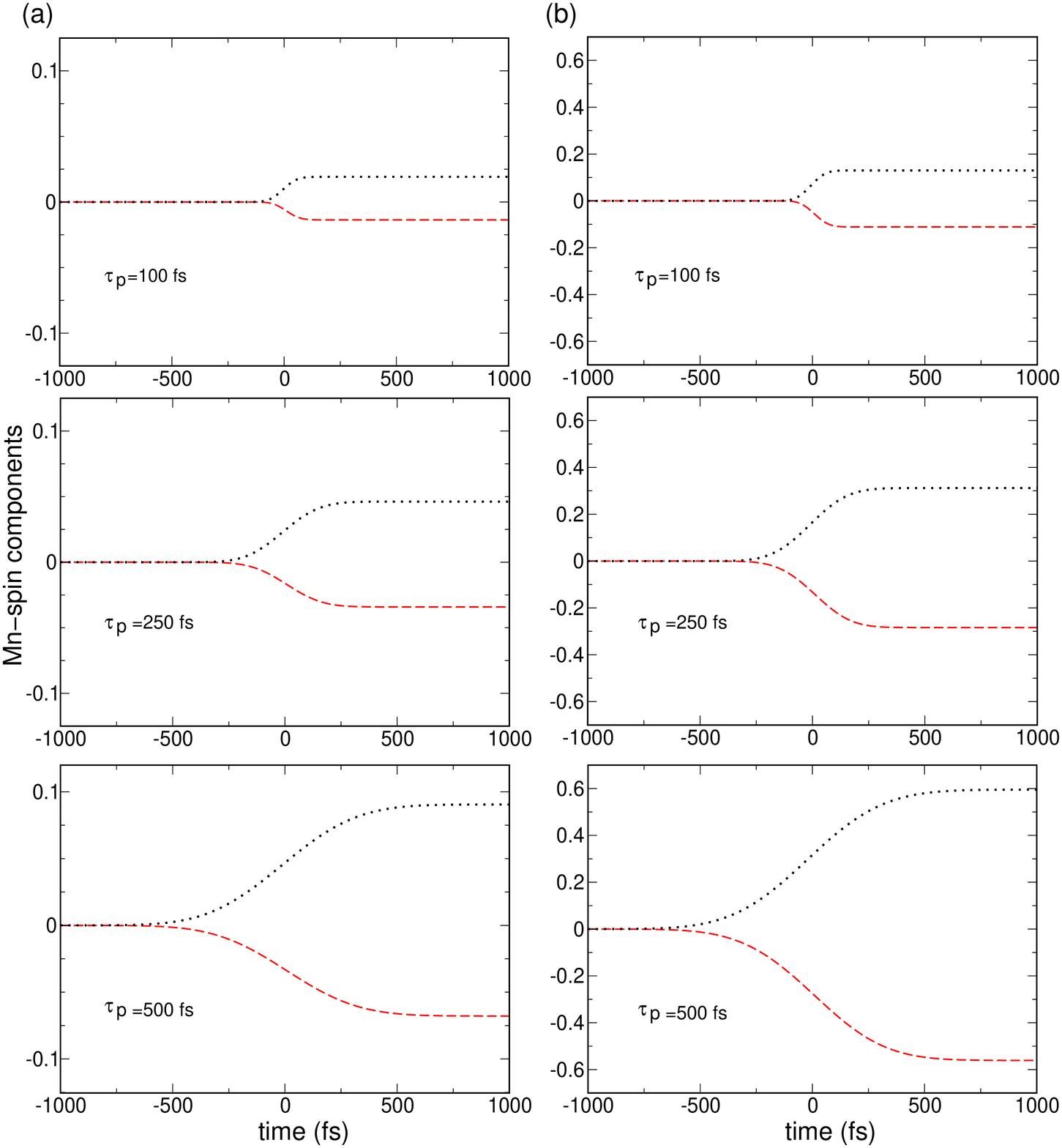,width=8.7cm}}
}
\caption{ (Color online) 
Mn spin components  without magnetic anisotropy 
for  pulse durations  and 
Rabi energies as in Fig.\ref{Fig1}.
$1/\Gamma_{\perp}$=21fs, $T_2$=10.5fs. 
Dashed line: ${\bf S}_y(t)$.
Dotted line: ${\bf S}_z(t)$
\label{Fig2}}
\end{figure}

\begin{figure}
\centerline{
\hbox{\psfig{figure=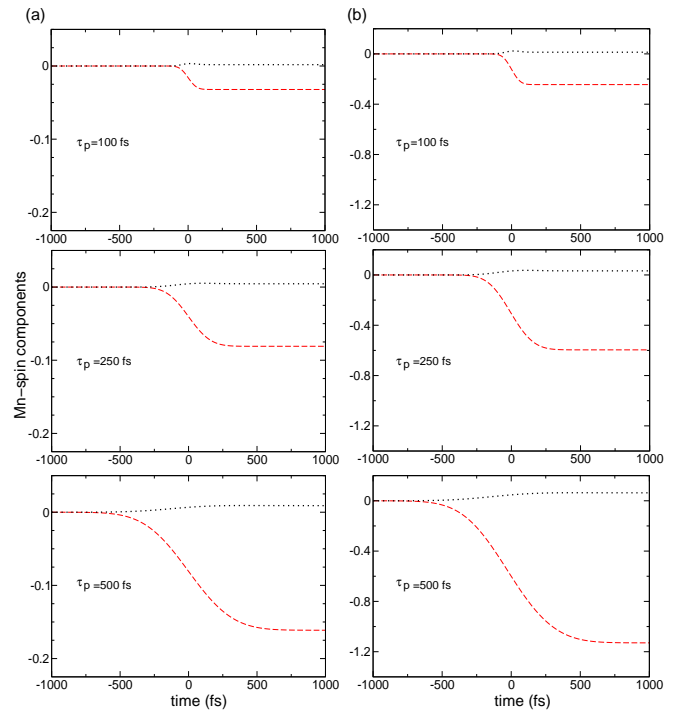,width=8.7cm}}
}
\caption{ (Color online) 
Time evolution of Mn spin components 
for the parameters of Fig.\ref{Fig2} except 
for $\Gamma_{\perp}$=0, $T_2$=330fs. 
\label{Fig3}}
\end{figure}

The Mn spin response is seen more clearly in Fig.\ref{Fig2}, 
which shows the development with time of
the y- and z--components 
of ${\bf S}$. 
Both of these  components vanish in the ground state. They 
 develop on a timescale 
determined by the pulse duration and have comparable magnitudes
for sufficiently large $\Gamma_{\perp}$. 
The importance  of the hole spin dephasing 
 can be seen
by comparing Fig.\ref{Fig2}, obtained 
for very short spin and polarization dephasing times, 
 with Fig.\ref{Fig3}, 
obtained for $\Gamma_{\perp}=0$. 
The most striking  difference 
is the very small magnitude of the out-of--plane 
component $S_z(t)$ when the hole spin is conserved 
($\Gamma_{\perp}=0$). 
The hole  spin follows  the overall Mn spin 
and remains more or less antiparallel to ${\bf S}(t)$ at all times,  
with the exception 
of a small component 
perpendicular to ${\bf S}$
that triggers the  spin dynamics.

We now turn to the effects 
of the magnetic anisotropy on the spin dynamics. 
As discussed above, the magnetic anisotropy
of the thermal  carriers leads to preferred 
Mn spin directions that minimize the total Fermi sea energy 
$E_h({\bf S})$, 
Eq.(\ref{tot-en}).
Its effects on the Mn spin 
dynamics can be treated phenomenologically 
by adding the
magnetic field 
\begin{equation} 
{\bf H}(t) = - \frac{
\partial E_h({\bf \hat{S}})}{\partial {{\bf \hat{S}}}}
\label{H}
\end{equation} 
to Eq.(\ref{Mn-spin}). \cite{halperin,halp-rev}
Such 
an anisotropy field causes a  nonlinear rotation 
and relaxation of the Mn spin 
until it
aligns with the easy axis direction that minimizes $E_h({\bf \hat{S}})$,
so that ${\bf H}=0$. 
We would like to distinguish between the above effect, due to the thermal carriers, 
and the anisotropy effects on the photoexcitation. The latter 
can be treated  microscopically as discussed in 
Section III and determine the photoexcited  hole spin
(see e.g. Eq.(\ref{hole-spin})). 

\begin{figure}
\centerline{
\hbox{\psfig{figure=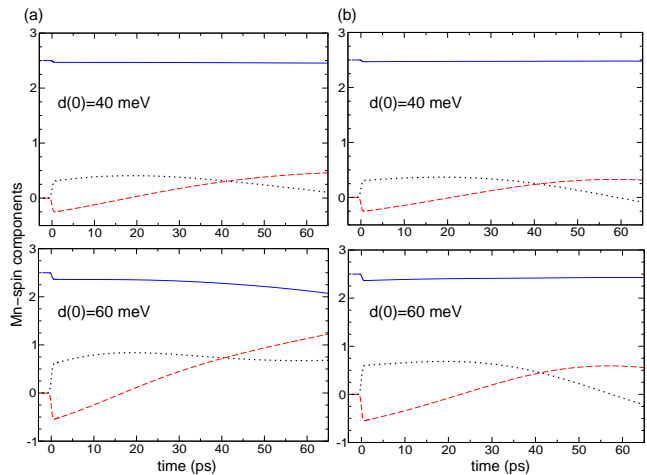,width=8.5cm}}
}
\caption{(Color online) 
Magnetic anisotropy effects on the 
Mn spin components for strong dephasing $1/\Gamma_{\perp}$=21fs, 
$T_2$=10.5fs, pulse duration $t_p$=500fs, 
and two Rabi energies. 
Solid line: ${\bf S}_x(t)$. 
Dashed line: ${\bf S}_y(t)$.
Dotted line: ${\bf S}_z(t)$.
 (a) and (b) compare the time evolution 
for initial condition along the two easy axes, 
 taken as the x--axis.
$K_c$=-0.0175meV, $K_{uz}$=0.0252meV, 
$K_u$=0.010meV.
\label{Fig4}}
\end{figure}

\begin{figure}
\centerline{
\hbox{\psfig{figure=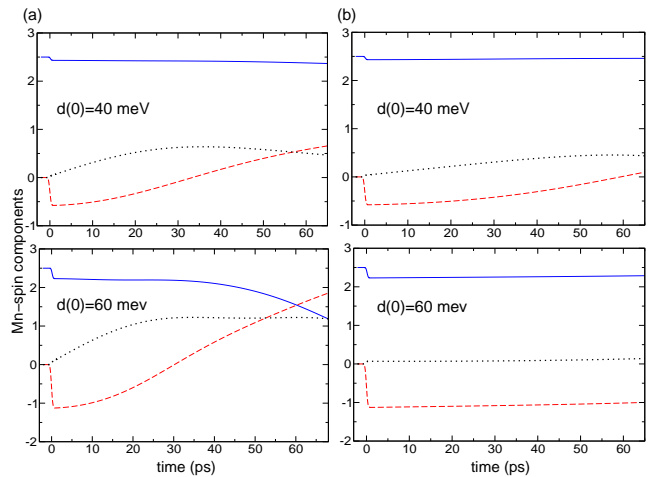,width=8.5cm}}
}
\caption{(Color online) 
Time evolution of Mn spin components 
for the parameters of Fig.\ref{Fig4} 
except for  $\Gamma_{\perp}$=0, 
$T_2$=330fs.
\label{Fig5}}
\end{figure}

In Ga(Mn)As, 
$K_c<0, K_u>0$, and $|K_c|>K_u$. \cite{welp,hamaya} 
In this parameter regime,
 there are two degenerate ground states, which 
correspond to Mn spin pointing 
at angles $\phi$ from the x--axis such that  
$\cos 2 \phi = K_u/|K_c|$.
The system can be prepared so that, prior to the photoexcitation, 
the Mn spin points along either one of the 
above two easy axes.   
In the absence of magnetic anisotropy, 
both these initial conditions would result in the same Mn spin 
trajectory. 
Figs.\ref{Fig4}(a) 
and \ref{Fig4}(b)  compare the three Mn spin components 
as function of time, starting with Mn spin pointing along 
the two different ground state configurations. 
To make this comparison more meaningful, 
in both cases we chose the x--axis of the coordinate system 
to coincide with the initial magnetization direction. 
Fig.\ref{Fig4} 
clearly shows that, as the Rabi energy increases, 
the two initial conditions 
lead to 
different time evolution, 
which implies that the two magnetic 
ground states can be distinguished.
As can be seen by 
 comparing Fig.\ref{Fig4}
with Fig.\ref{Fig5}, 
the Mn spin dynamics
depends on the hole spin dephasing,
especially as the photoexcitation intensity increases.

\begin{figure}
\centerline{
\hbox{\psfig{figure=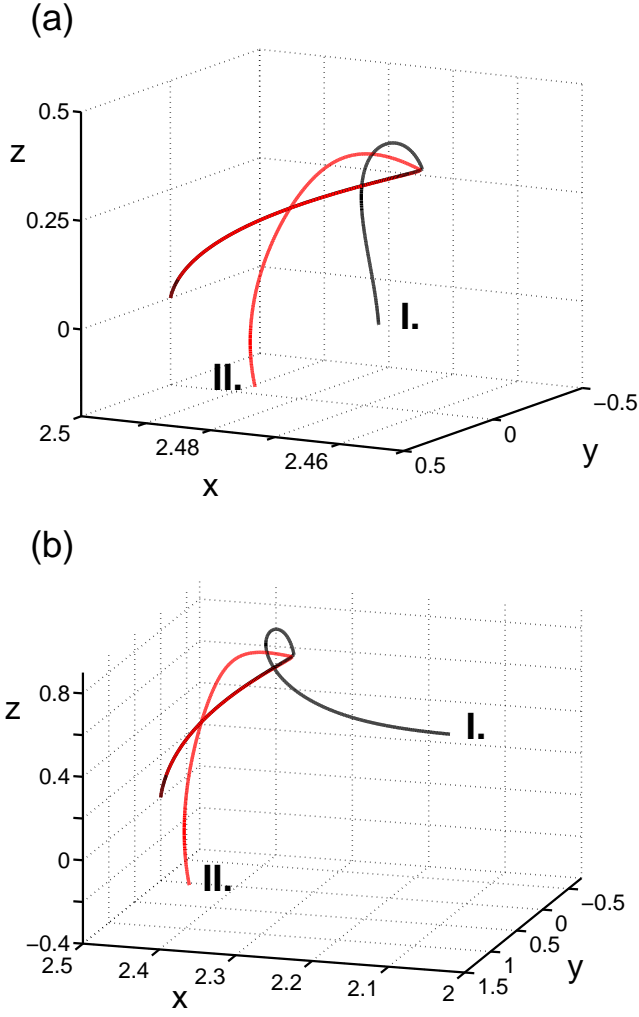,width=8.5cm}}
}
\caption{(Color online) 
Mn spin trajectories with
initial condition along the two different easy axes 
(I and II) for  
the parameters of Fig.\ref{Fig4}.  
(a): $d_+$=40meV, (b): $d_+$=60meV.}
\label{Fig6}
\end{figure}

\begin{figure}
\centerline{
\hbox{\psfig{figure=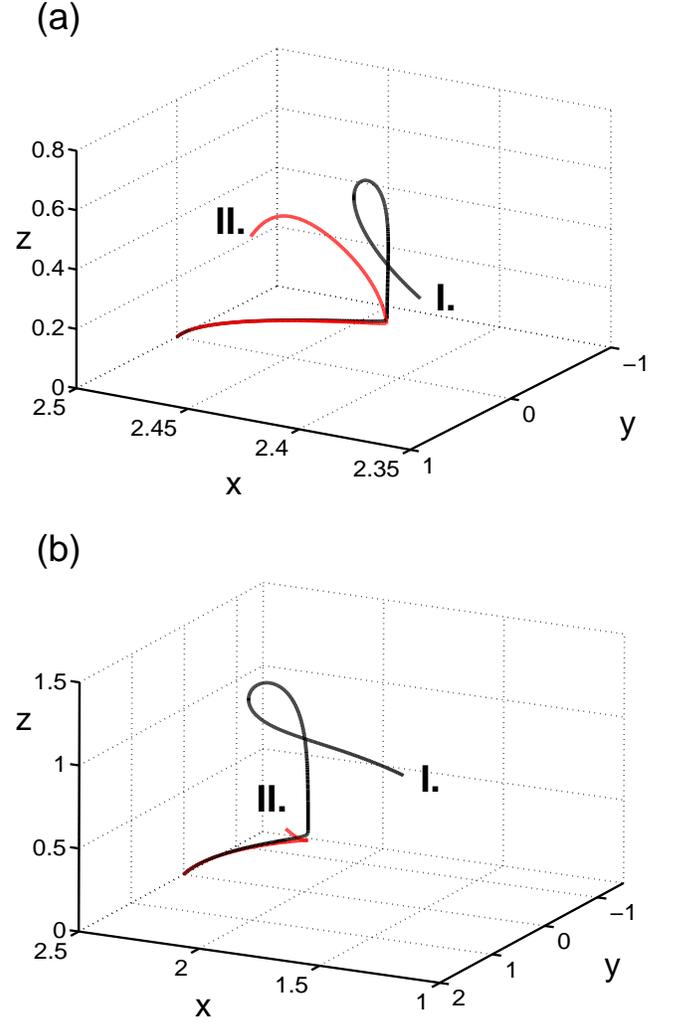,width=8.5cm}}
}
\caption{(Color online) 
Mn spin trajectories 
for the same parameters as in Fig.\ref{Fig6} 
except for $\Gamma_{\perp}$=0, 
$T_2$=330fs.
\label{Fig7}}
\end{figure}

The dependence of the Mn spin dynamics 
on the  initial magnetic state
becomes more clear in Figs.\ref{Fig6} and 
\ref{Fig7}, 
which plot
the Mn spin trajectories for the two initial conditions 
in the same coordinate system, 
whose x--axis coincides
with the initial magnetizations.
Two different temporal regimes 
can be clearly distinguished. 
The first regime lasts for a time interval comparable to the pulse 
duration. Here the dynamics
is governed by the optical pulse intensity, duration, and helicity, 
as well as by the ratio $\mu_+/\mu_-$
and the mixing of the hole spin states
that depend on the magnetic anisotropy.  
On the other hand,
the magnetic anisotropy field Eq.(\ref{H}), due to the 
thermal carriers,  
plays a very minor role
in this initial temporal regime  for 
sufficiently large Rabi energies that exceed 
the anisotropy constants.
The  trajectories for the different initial conditions 
coincide in this sub--picosecond regime, which 
gives an 
ultrafast  magnetization
tilt 
determined by the response 
to the hole spin  photoexcited 
by the coherent spin Raman processes 
described by Eqs.(\ref{hole-spin})
and (\ref{h}).  
As can be seen by comparing 
Figs.\ref{Fig6} and 
\ref{Fig7}, the shape of the Mn spin trajectories is strongly 
influenced by the spin and polarization dephasing.

Figs.\ref{Fig6} and 
\ref{Fig7} show a rather abrupt change in the 
Mn spin trajectory at times $t \sim \tau_p$. 
At later times, the temporal evolution is determined by the magnetic 
anisotropy field Eq.(\ref{H}).
The shape of the trajectory during many picoseconds 
depends critically on the magnetization tilt 
that develops during the femtosecond initial stage. 
If this tilt is sufficiently small, the Mn spin dynamics is 
described by linearized equations of motion, expanded 
around the equilibrium spin values that minimize the Fermi sea magnetic energy 
$E_h({\bf S})$. This harmonic oscillation 
 corresponds to zero--momentum magnon 
excitations, whose frequency is the same 
for the two easy axes. 
Since the magnitude of the magnetization tilt 
can be controlled by the optical pulse intensity and duration, 
we conclude that for sufficiently short pulses and 
sufficiently small Rabi energies, $d_{+} \tau_p < 1$, 
the Mn spin trajectories are very similar for both initial conditions. 
They correspond to magnetization precession 
around the easy axis (magnons).  
However, with increasing photoexcitation intensity 
and duration, 
the deviation 
from the easy axis due to the sub--picosecond magnetization tilt 
increases, and eventually  
the Mn spin dynamics cannot be described by expanding 
$E_h({\bf S})$ around its minimum. Such nonlinearities  result
in a complex trajectory, determined by the full nonlinear 
equations Eqs.(\ref{Mn-spin}) and (\ref{H}), which can differ substantially 
from a simple magnon precession.
As demsonstrated by  Figs.\ref{Fig6} and 
\ref{Fig7}, in this nonlinear dynamics regime, 
the Mn spin trajectories depend on the initial 
condition. 
One can then distinguish 
between the different magnetic states,
which 
can provide the basis for an ultrafast 
magnetic memory read--out scheme with speed 
 limited only by the optical pulse duration. 
Such a  scheme would be based on controlling the ultrafast response of the  
 spin system 
via the optical pulse intensity, duration,
and helicity. 
Furthermore, the strong dependence of the trajectory 
shape on the spin dephasing implies that the valence bandstructure 
and mixing of the hole spin states by the spin orbit interaction
plays an important role.
In the next section we interpret our numerical results
by deriving from the full theory 
an effective Landau--Gilbert--like equation for the Mn spin
after expanding around the adiabatic limit.

\section{Interpretation: light-induced relaxation and 
re-orientation} 
\label{sec:inter}

The  numerical solution of the full mean field 
equations, discussed in the previous section,
shows that 
the hole spin 
follows the Mn spin more or less   adiabatically. This is due to 
the much faster  hole spin precession 
and relaxation as compared 
to the time--dependent changes in ${\bf S}(t)$.
To interpret the full numerical results, 
we therefore 
expand around the adiabatic limit.
The  derivation presented in Appendix \ref{adiabatic} 
then gives 
a Landau--Gilbert-like \cite{landau} 
 equation:
\begin{eqnarray} 
 \partial_t {\bf \hat{S}} &=&
( 1 - \gamma  s^{h}_{ \parallel})
{\bf \hat{S}} \times {\bf H} 
+ \alpha  s^{h}_{ \parallel}
{\bf \hat{S}} \times \left( {\bf \hat{S}} \times {\bf H} \right)
\nonumber \\
&-& \alpha {\bf \hat{S}} \times {\bf h}(t) 
- \gamma {\bf \hat{S}} \times \left( {\bf \hat{S}} \times {\bf h}(t)\right), 
\label{eom-eff} 
\end{eqnarray} 
where the effective magnetic fields ${\bf h}$
(Eq.(\ref{h})), 
\begin{equation} 
{\bf h}(t)=\frac{1}{V} \sum_{{\bf k}} \ Im \ {\bf h}_{{\bf k}}(t),
\label{h-eff} 
\end{equation} 
and ${\bf H}$, Eq.(\ref{H}), are due to 
the optical excitation and magnetic anisotropy 
respectively.
In the above equation, 
\begin{equation} 
\gamma=\frac{\beta  \left( \beta c S +  
\beta s^{h}_{ \parallel} +  H_{\parallel}\right)}{
\left( \beta c S +  \beta s^{h}_{ \parallel} +  
H_{\parallel}\right)^2 + \Gamma_{\perp}^2}
\label{gamma} 
\end{equation} 
where $H_{\parallel} = {\bf H} \cdot {\bf \hat{S}}$, while  
\begin{equation} 
\alpha = 
 \frac{ \Gamma_{\perp} \beta }{ 
\left( \beta c S +  \beta s^{h}_{ \parallel} +  
H_{\parallel}\right)^2 + \Gamma_{\perp}^2}.
\label{Gilbert} 
\end{equation}
The first two terms on the rhs of Eq.(\ref{eom-eff}) 
correspond to the 
usual Landau--Gilbert description of the 
 Mn spin dynamics induced by 
the magnetic anisotropy
field ${\bf H}$. They are  similar to the results 
of Refs. \onlinecite{halperin,mitchell,sinova} 
and also include nonlinear corrections.   
The first term on the rhs 
describes 
the Mn spin rotation around 
the anisotropy field ${\bf H}$. 
$\gamma$, Eq.(\ref{gamma}), gives  
the renormalization of the gyromagnetic ratio
by the magnetic exchange interaction. 
The second term 
describes the Gilbert  relaxation 
of the Mn spin towards the direction of ${\bf H}$, 
with Gilbert damping coefficient $\propto \alpha$, Eq.(\ref{Gilbert}). 
The precise values of $\gamma$ and $\alpha$ 
in the realistic system 
depend on the details of the valence bandstructure.
The relaxation of the 
hole spin component 
parallel to ${\bf S}(t)$, $s^{h}_{ \parallel}$,  
depends on 
the two fields ${\bf h}(t)$ (nonlinear optical excitation) and ${\bf H}(t)$
(magnetic anisotropy)  
as described in Appendix \ref{parallel}.

The nonlinear optical excitation gives rise to 
two additional contributions to
the Mn spin equation of motion Eq.(\ref{eom-eff}). 
These contributions 
describe the response of the magnetization 
to the effective magnetic field pulse ${\bf h}(t)$, which is
 generated  by the optical excitation 
via second order processes. 
The third term on the rhs of  Eq.(\ref{eom-eff}) 
describes a Mn spin rotation around 
${\bf h}(t)$, whose magnitude is 
proportional to the 
Gilbert damping coefficient $\alpha$, Eq.(\ref{Gilbert}),  and 
therefore  vanishes if $\Gamma_{\perp}=0$ 
(hole spin  conserved). 
The last term describes a  spin relaxation 
towards 
${\bf h}(t)$, determined by the exchange interaction. 
The above light--induced rotation and relaxation vanish after 
the decay of the optical pulse and e--h polarizations. 

\begin{figure}
\centerline{
\hbox{\psfig{figure=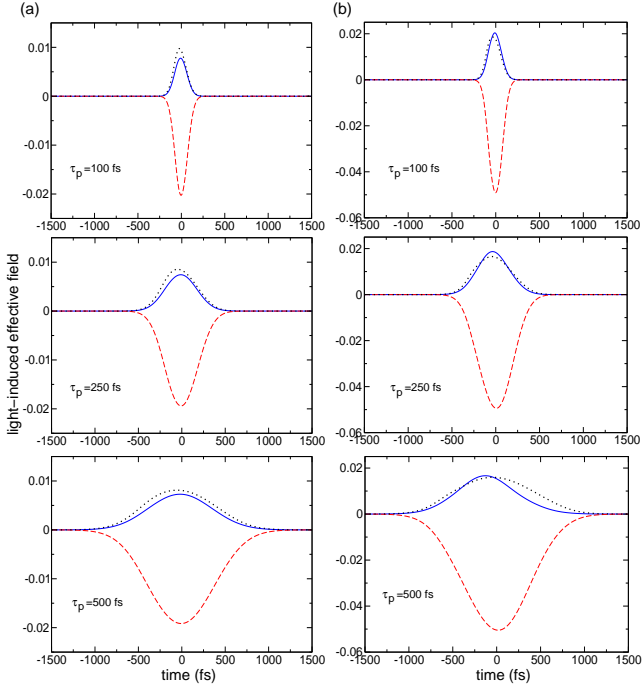,width=8.5cm}}
}
\caption{(Color online) 
Light--induced effective magnetic field 
components 
for  pulse durations  and 
Rabi energies as in Fig.\ref{Fig1}.
$1/\Gamma_{\perp}$=21fs, $T_2$=10.5fs. 
Solid line: ${\bf h}_x(t)$, 
Dashed line: ${\bf h}_y(t)$,
Dotted line: ${\bf h}_z(t)$. 
\label{Fig8}}
\end{figure}

\begin{figure}
\centerline{
\hbox{\psfig{figure=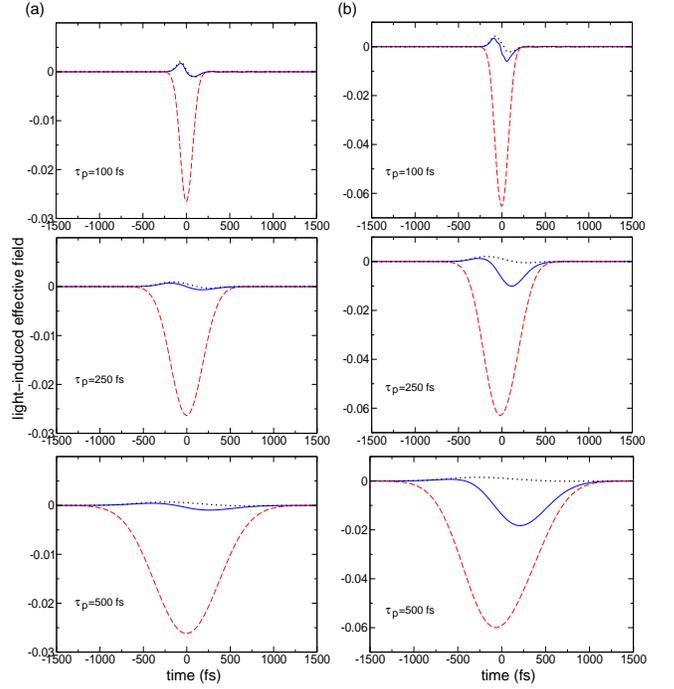,width=8.5cm}}
}
\caption{(Color online) 
Same as Fig.\ref{Fig8} 
but with $\Gamma_{\perp}$=0, 
$T_2$=330fs.
\label{Fig9}}
\end{figure}

Eq.(\ref{eom-eff}) predicts an 
extremely fast response of the spin system, determined by the effective 
magnetic field pulse
${\bf h}(t)$,  
whose speed is limited 
only by the pulse duration.  This result 
points to a new way of controlling the magnetization
during femtosecond time scales, much shorter than
those accomplished  so far.
The last two terms on the rhs of Eq.(\ref{eom-eff}) 
demonstrate 
an ultrafast 
modification of the effective magnetic field ${\bf H}$. 
Eq.(\ref{eom-eff}) therefore predicts 
 two regimes of time evolution, 
which can be separated 
due to their very different timescales. 
The initial temporal regime is controlled by the pulsed field ${\bf h}(t)$ 
and vanishes soon after the optical excitation. The second 
temporal regime is controlled by the anisotropy field 
${\bf H}(t)$, 
 Eq.(\ref{H}), due to the thermal Fermi sea, which depends nonlinearly on ${\bf S}(t)$. 

We now turn to the effective light--induced magnetic field pulse ${\bf h}(t)$.
 As can be seen from 
Eq.(\ref{eom-eff}), in order to trigger  magnetization dynamics, 
 ${\bf h}(t)$ must have a component perpendicular to the ground state 
Mn spin and the easy axis.  
The direction of ${\bf h}(t)$ 
is determined by the optical transition selection rules and 
by the interband polarizations.
Figs.\ref{Fig8} and \ref{Fig9} 
show the three components 
${\bf h}_x(t)$, ${\bf h}_y(t)$, and ${\bf h}_z(t)$
in the case of strong spin dephasing (Fig.\ref{Fig8}) or
uncoupled hole spin states (Fig.\ref{Fig9}). 
It is clear by comparing the two figures that 
the direction of ${\bf h}(t)$, 
which governs the Mn spin trajectory,
depends  on the dephasing. 
The mosty striking effect
of $\Gamma_{\perp}$ 
is the development, on the time scale of the optical excitation, 
of a large component ${\bf h}_z$ in the direction of optical pulse propagation. 
This component  
is strongly suppressed in the case of weak dephasing. 
As can be seen from 
Eq.(\ref{eom-eff}), ${\bf h}_z$ rotates the Mn spin 
within the x--y plane, in a counter--clockwise direction,   
and leads to spin relaxation towards 
the positive z--axis, i.e. to a  magnetization tilt out of the x--y plane. 
On the other hand, for $\Gamma_{\perp}=0$, 
the Gilbert damping coefficient vanishes, $\alpha=0$, 
which suppresses the light--induced Mn spin rotation.
The only effect of the photoexcitation is then the 
relaxation described by the last term 
of Eq.(\ref{eom-eff}), which causes the Mn spin to relax towards 
the negative y--axis (counter--clockwise rotation within the 
$x-y$ plane) since ${\bf h}_z \approx 0$ in this case.  
The above trends  are consistent with the full numerical calculation
of the mean field equations discussed in the previous section.  
In Appendix \ref{h-calc}
we derive an equation of motion 
for ${\bf h}(t)$ that  demonstrates its dependence on the 
optical transition selection rules, the hole spin polarization,     
the magnetic exchange interaction, and the hole spin dephasing.

\section{Conclusions} 
\label{sec:concl} 

In conclusion, we presented a theory of 
nonlinear spin dynamics in 
ferromagnetic semiconductors and demonstrated the main effects 
common in all such systems
by considering a single spin--degenerate electron and hole band. 
Our main result of relevance to experiments is the existence of two different temporal regimes 
of magnetization dynamics. The initial regime lasts for 
time scales comparable to the optical pulse duration. It is 
governed by magnetization precession and relaxation 
around an effective magnetic field pulse generated by the 
optical excitation via 
second--order coherent nonlinear processes. 
 The second temporal regime is governed by the magnetic 
anisotropy due to the thermal carriers. 
We showed that the shape of the magnetization trajectory 
depends sensitively on the hole spin dephasing, 
the magnetic easy axes, the transition matrix elements, 
 and the optical 
pulse intensity, duration, and helicity. 
The latter determine the magnitude and direction 
of the ultrafast magnetization tilt 
from the magnetic easy axis 
in response to the photoexcitation of hole spin, 
which in turn determines the importance of nonlinear magnetic effects
during many picoseconds.  
To interpret our numerical results, 
we derived an effective Landau--Gilbert--like equation of motion.  
This equation shows that 
an effective magnetic field pulse, generated by  the interband optical polarizations
via a Raman--like process, 
triggers precession and relaxation during femtosecond time scales.
This magnetic field pulse may be thought of as an ultrafast correction 
to the anisotropy field. 
The  picture of ultrafast magnetization dynamics conveyed by 
our  results and, in particular, the existence of two temporal regimes 
is consistent with the recent experimental observation 
of a light--induced  modification of the magnetic anisotropy field
in Ga(Mn)As during femtosecond timesales, induced by nonthermal high energy carriers.
\cite{wang-08} A more detailed comparison between theory and experiment 
must be performed in the future.  
Our calculations  point out the need for further experiments in order to explore 
the control of the magnetization trajectory by changing the 
optical excitation intensity, duration, central frequency, and helicity. 
For a complete understanding of the 
magnetization dynamics in   ferromagnetic semiconductors,  
the role of fluctuations and carrier--spin correlations, \cite{kapet}
the bandstructure, and disorder must also be considered.

This work was supported by the EU 
STREP program HYSWITCH.

%****

\appendix 
\section{}
\label{spin-deph} 

In this Appendix we derive the relaxation and dephasing contributions to the 
hole spin equations of motion. 
The hole spin is expressed in the 
basis of hole spin eigenstates as 
 \begin{eqnarray} 
{\bf s}_{{\bf k} m z}^h = \frac{1}{2} 
\left[ \langle \hat{h}_{-{\bf k} m \uparrow}^\dag \hat{h}_{-{\bf k} m \uparrow}
\rangle - 
\langle \hat{h}_{-{\bf k} m \downarrow}^\dag \hat{h}_{-{\bf k} m \downarrow}
\rangle \right], \label{def-hole-spin-z}\\
{\bf s}_{{\bf k}m +}^h = 
 {\bf s}_{{\bf k} m x}^h + i  
{\bf s}_{ {\bf k} m y}^h = 
\langle \hat{h}_{-{\bf k}m \uparrow}^\dag \hat{h}_{-{\bf k} m \downarrow}
\rangle. 
\label{def-hole-spin-perp}
\end{eqnarray} 
From Eqs.(\ref{lind}) and (\ref{lind-op})  we 
 obtain for the corresponding hole density matrices 
after using straightforward algebra and the property 
$\hat{h}^2 =0$:
\begin{eqnarray} 
\partial_t \left. \langle 
\hat{h}^{\dag}_{ -{\bf k}m \Downarrow}
\hat{h}_{-{\bf k}m\Uparrow} \rangle
\right|_{relax}= - \frac{\Gamma_\perp}{2} 
\langle \hat{h}^{\dag}_{ -{\bf k}m \Downarrow}
\hat{h}_{-{\bf k} m \Uparrow} \rangle
\label{relax-h-perp}
\end{eqnarray} 
By choosing the z--axis parallel to the Mn spin 
and noting the expression 
Eq.(\ref{def-hole-spin-perp}),
we obtain 
Eq.(\ref{spineom-perp}) 
that describes the dephasing of the hole spin component 
${\bf s}^{h}_{{\bf k}m{\perp}}$
perpendicular to the Mn spin. 
Similarly we obtain from  Eqs.(\ref{lind}) and (\ref{lind-op})  
\begin{eqnarray} 
&&\partial_t \left. \langle \hat{h}^{\dag}_{ -{\bf k}m \Downarrow}
\hat{h}_{-{\bf k} m \Downarrow} \rangle
\right|_{relax}=  - 
\partial_t \left. \langle \hat{h}^{\dag}_{-{\bf k} m \Uparrow}
\hat{h}_{-{\bf k} m \Uparrow} \rangle
\right|_{relax} \nonumber \\
&& = \Gamma_\perp 
\Bigg \lbrack \langle \hat{h}^{\dag}_{ -{\bf k} m \Uparrow}
\hat{h}_{-{\bf k} m \Uparrow} \rangle
\nonumber \\
&&
- \langle \hat{h}^{\dag}_{ -{\bf k} m \Uparrow}
\hat{h}^{\dag}_{ -{\bf k} m \Downarrow}
\hat{h}_{-{\bf k} m \Downarrow}
\hat{h}_{-{\bf k} m \Uparrow} \rangle
\Bigg \rbrack,
\label{relax-h-par} 
\end{eqnarray}
which  conserve the total hole population
\begin{equation} 
N^h_{{\bf k} m} 
=  \langle \hat{h}^{\dag}_{-{\bf k} m \Downarrow}
\hat{h}_{-{\bf k} m \Downarrow} \rangle + \langle \hat{h}^{\dag}_{
-{\bf k} m \Uparrow}
\hat{h}_{-{\bf k} m \Uparrow} \rangle.
\label{N}
\end{equation} 
Using the factorization 
\begin{eqnarray} 
&&\langle \hat{h}^\dag_{1} \hat{h}^\dag_2
\hat{h}_{3} \hat{h}_4 \rangle 
= \langle \hat{h}^\dag_{1} 
\hat{h}_{4}\rangle  
\langle \hat{h}^\dag_2 \hat{h}_3 \rangle
- \langle \hat{h}^\dag_{1} 
\hat{h}_{3}\rangle  
\langle \hat{h}^\dag_2 \hat{h}_4 \rangle
\end{eqnarray} 
and the relation
(obtained 
from  Eqs.(\ref{def-hole-spin-z})
and (\ref{N}))
\begin{eqnarray} 
\langle \hat{h}_{-{\bf k} m \sigma}^\dag \hat{h}_{-{\bf k} m \sigma}
\rangle 
=  N^h_{{\bf k} m }/2 + \sigma s^{h}_{{\bf k}m \parallel},
\label{n-N} 
\end{eqnarray}
where $\sigma=\pm 1$, 
 we obtain Eq.(\ref{spineom-par}) 
for the 
hole spin relaxation.

\section{} 
\label{pol-deph} 

In this appendix we derive the 
polarization dephasing 
$\left. \partial_t P_{{\bf k} m n \sigma} \right|_{relax}$
by using Eqs.(\ref{lind}) and (\ref{lind-op}). 
After 
straightforward algebra
and using the 
properties 
$L^{\dag}_{{\bf k} m} \hat{h}_{{\bf k} m \sigma} L_{{\bf k} m} = 0$,  
$\hat{h}^2=0$, we obtain 
after noting that the terms 
in the summation of Eq.(\ref{lind}) vanish  unless 
 $({\bf k}^\prime m^\prime) = ({\bf k} m)$:   
\begin{eqnarray} 
\left. \partial_t P_{{\bf k} m  n \sigma} \right|_{relax} 
= - \frac{\Gamma_\perp}{2} \langle  
\hat{h}^{\dag}_{-{\bf k} m \Uparrow} \hat{h}_{-{\bf k} m \Uparrow} 
\hat{h}_{-{\bf k} m \sigma} \hat{e}_{{\bf k}n}\rangle 
\nonumber \\
- \frac{\Gamma_\perp}{2} \langle   
\hat{h}_{-{\bf k}m \sigma} \hat{e}_{{\bf k} n}  
\hat{h}_{-{\bf k}m \Downarrow} \hat{h}^{\dag}_{-{\bf k}m \Downarrow} \rangle.
\label{step1} 
\end{eqnarray}  
We now note the relation between 
the hole operators  with spin 
along the z--axis of pulse propagation or along the 
Mn spin direction: 
\begin{eqnarray} 
\hat{h}_{{\bf k} m\Uparrow}
= e^{i \phi/2} \, \cos\frac{\theta}{2} 
\, \hat{h}_{{\bf k} m \uparrow}
+ e^{-i \phi/2} \, \sin\frac{\theta}{2} \,  
\hat{h}_{{\bf k} m \downarrow}, \label{spin-up} \\ 
\hat{h}_{{\bf k} m \Downarrow}
= - e^{i \phi/2} \, \sin \frac{\theta}{2} \, 
\hat{h}_{{\bf k} m \uparrow}
+ e^{-i \phi/2} \, \cos\frac{\theta}{2} \, 
\hat{h}_{{\bf k} m \downarrow} \label{spin-down} 
\end{eqnarray} 
where  $\theta(t)$ and $\phi(t)$ are the polar coordinates 
that define the direction of  the Mn spin ${\bf S}(t)$
and the operators 
$\hat{h}_{{\bf k} m \uparrow}$ 
and $\hat{h}_{{\bf k} m \downarrow}$ 
create hole states with spin along the z--axis.
Using Eqs. (\ref{spin-up}) 
and (\ref{spin-down}), 
the property $\hat{h}^2=0$,
and  the relations 
${\bf S}^+/S=e^{i \phi} \sin \theta,  {\bf S}_z/S =\cos \theta$
we obtain from 
Eq.(\ref{step1}) 
 after straightforward algebra 
\begin{eqnarray} 
&&\left. \partial_t P_{{\bf k} m n \uparrow} \right|_{relax} 
 = \nonumber \\
&&- \Gamma_\perp 
\left[\frac{ 1 - {\bf S}_z/S}{2}  \langle  
\hat{h}^{\dag}_{-{\bf k} m \downarrow} \hat{h}_{-{\bf k} m \downarrow} 
\hat{h}_{-{\bf k}m \uparrow} 
\hat{e}_{{\bf k} n \downarrow}\rangle \right. \nonumber \\
&&\left. + \frac{{\bf S}_{-}}{2S} 
\langle  \hat{h}^\dag_{-{\bf k} m \uparrow} \hat{h}_{-{\bf k} m \downarrow}
\hat{h}_{-{\bf k} m \uparrow} 
\hat{e}_{{\bf k}n \downarrow}\rangle \right]  \nonumber \\
&&- \Gamma_\perp
\left[ \frac{ 1 + {\bf S}_z/S}{2} 
\langle  \hat{h}_{-{\bf k} m \uparrow} 
\hat{e}_{{\bf k} n \downarrow}
\hat{h}_{-{\bf k} m \downarrow} \hat{h}^\dag_{-{\bf k} m \downarrow} \rangle 
\right. \nonumber \\
&& \left. - \frac{{\bf S}_{-}}{2S} \langle 
\hat{h}_{-{\bf k} m \uparrow} 
\hat{e}_{{\bf k} n \downarrow}
 \hat{h}_{-{\bf k} m \downarrow}
 \hat{h}^\dag_{-{\bf k} m \uparrow} \rangle  
\right]
\end{eqnarray} 
and 
\begin{eqnarray} 
&&\left. \partial_t P_{{\bf k} m n \downarrow} \right|_{relax} 
 =  \nonumber \\
&&- \Gamma_{\perp} 
\left[\frac{ 1 + {\bf S}_z/S}{2}  \langle  
\hat{h}^{\dag}_{-{\bf k} m \uparrow} \hat{h}_{-{\bf k} m \uparrow} 
\hat{h}_{-{\bf k} m \downarrow} 
\hat{e}_{{\bf k} n \downarrow}\rangle \right. \nonumber \\
&&+ \left. \frac{{\bf S}_{+}}{2S} 
\langle  \hat{h}^\dag_{-{\bf k} m \downarrow} \hat{h}_{-{\bf k}m \uparrow}
\hat{h}_{-{\bf k} m \downarrow} 
\hat{e}_{{\bf k} n \downarrow}\rangle \right] \nonumber \\
&&- \Gamma_{\perp}
\left[ \frac{ 1 - {\bf S}_z/S}{2} \langle 
\hat{h}_{-{\bf k} m \downarrow} 
\hat{e}_{{\bf k} n \downarrow}
\hat{h}_{-{\bf k} m \uparrow} \hat{h}^\dag_{-{\bf k} m \uparrow} 
\rangle \right. \nonumber \\
&& - \frac{{\bf S}_{+}}{2S}  \langle  \left. 
\hat{h}_{-{\bf k} m \downarrow} 
\hat{e}_{{\bf k} n \downarrow}
\hat{h}_{-{\bf k} m \uparrow} \hat{h}^\dag_{-{\bf k} m \downarrow} \rangle 
\right]. 
\end{eqnarray}
We now factorize the higher density matrices in the above 
equations as follows: 
\begin{eqnarray}
\langle \hat{h}^\dag_1 \hat{h}_2 \hat{h}_3 \hat{e} \rangle = 
\langle \hat{h}^\dag_1 \hat{h}_2 \rangle \langle \hat{h}_3 
\hat{e} \rangle  
- \langle \hat{h}^\dag_1 \hat{h}_3 \rangle 
\langle  \hat{h}_2 \hat{e} \rangle, \\
\langle \hat{h}_3 \hat{e} \hat{h}_2  \hat{h}^\dag_1 \rangle 
=  \langle \hat{h}_3 \hat{e} \rangle \langle 
\hat{h}_2  \hat{h}^\dag_1\rangle 
-\langle \hat{h}_2 \hat{e} \rangle \langle 
\hat{h}_3  \hat{h}^\dag_1 \rangle. 
\end{eqnarray} 
We thus obtain that
\begin{eqnarray} 
&&\left. \partial_t P_{{\bf k} m n \uparrow} \right|_{relax} 
 = \nonumber \\
&&
 - \Gamma_\perp 
\left\{\frac{ 1 - {\bf S}_z/S}{2} 
\left( n^h_{{\bf k} m \downarrow} 
P_{{\bf k} m n \uparrow} 
- {\bf s}^h_{{\bf k} m -} 
P_{{\bf k} m n \downarrow} \right) \right. \nonumber \\
 && + \frac{{\bf S}_{-}}{2S} \left( {\bf s}^h_{{\bf k} m +}
P_{{\bf k} m n \uparrow} 
- n^h_{{\bf k} m \uparrow} 
P_{{\bf k} m n \downarrow} \right) 
\nonumber \\
&& + \frac{ 1 + {\bf S}_z/S}{2} 
\left[ P_{{\bf k} m n \uparrow}
( 1 - n^h_{{\bf k} m \downarrow}) 
+ 
P_{{\bf k} m n  \downarrow}
{\bf s}^h_{{\bf k} m -} \right] \nonumber \\
&& \left. + \frac{{\bf S}_{-}}{2S}
\left[ {\bf s}^h_{{\bf k} m +} 
P_{{\bf k} m n \uparrow}
 + 
P_{{\bf k} m n \downarrow}
( 1 - n^h_{{\bf k} m \uparrow}) \right] 
\right\}.
\end{eqnarray} 
and 
\begin{eqnarray} 
&&\left. \partial_t P_{{\bf k} m n \downarrow} \right|_{relax} 
 = \nonumber \\
&& - \Gamma_\perp 
\left\{\frac{ 1 + {\bf S}_z/S}{2} 
\left( n^h_{{\bf k} m \uparrow} 
P_{{\bf k} m n \downarrow} 
- {\bf s}^h_{{\bf k} m +} 
P_{{\bf k} m n \uparrow} \right) \right. \nonumber \\
&&  + \frac{{\bf S}_{+}}{2S} \left( {\bf s}^h_{{\bf k} m -}
P_{{\bf k} m n \downarrow} 
- n^h_{{\bf k} m \downarrow} 
P_{{\bf k} m n \uparrow} \right) 
\nonumber \\
 &&+ \frac{ 1 - {\bf S}_z/S}{2} 
\left[ P_{{\bf k} m n \downarrow}
( 1 - n^h_{{\bf k} m \uparrow}) 
+ 
P_{{\bf k} m n \uparrow}
{\bf s}^h_{{\bf k} m +} \right] \nonumber \\
&&\left. + \frac{{\bf S}_{+}}{2S}
\left[ {\bf s}^h_{{\bf k} m -} 
P_{{\bf k} m n \downarrow}
 + 
P_{{\bf k} m n \uparrow}
( 1 - n^h_{{\bf k} m \downarrow}) \right] 
\right\},
\end{eqnarray} 
where $n_{{\bf k} m \sigma}$ are the spin--polarized hole populations. 
After expressing the latter by using 
Eq.(\ref{n-N}) we obtain after some algebra 
the  polarization dephasing contributions 
Eqs. (\ref{pol-deph-up}) 
and (\ref{pol-deph-down}).

\section{} 
\label{adiabatic} 

In this appendix we derive an effectve 
equation of motion for the Mn spin 
by expanding around the adiabatic limit. 
First 
we use the hole spin decomposition  
Eq.(\ref{decomp}) 
into a component parallel   ${\bf \hat{S}}(t)$
($s^{h}_{{\bf k} \parallel}$) 
and a component perpendicular to ${\bf \hat{S}}(t)$
(${\bf s}^{h}_{{\bf k}{\perp}}$). 
The  unadiabatic contribution ${\bf s}^{h}_{{\bf k}{\perp}}$
is responsible for triggering the 
Mn spin precession. 
After substituting Eq.(\ref{decomp})  into Eq.(\ref{hole-spin}) 
and noting that the Mn spin magnitude 
$S$ remains constant in time we obtain 
\begin{eqnarray} 
\partial_t  {\bf s}^{h}_{{\bf k}{\perp}}
=\beta c{\bf S}\times{\bf s}^h_{{\bf k} \perp}+
\, \ Im \ {\bf h}_{{\bf k}}(t)
-  s^{h}_{{\bf k} \parallel} \partial_t {\bf \hat{S}} 
\nonumber \\
- {\bf \hat{S}} \left[ \partial_t 
s^{h}_{{\bf k} \parallel}
+ \Gamma_{\parallel} 
\left(
s^{h}_{{\bf k} \parallel}
+ m^h_{{\bf k}} \right) \right]
- \Gamma_{\perp}  
{\bf s}^{h}_{{\bf k} \perp}.
\end{eqnarray} 
By projecting out the components  parallel 
and perpendicular to ${\bf \hat{S}}$ 
and using the relations 
${\bf S} \cdot \partial_t {\bf S}
=  \partial_t S^2/2 =0$
and 
${\bf s}^{h}_{{\bf k} \perp} \cdot {\bf S}=0$
we obtain that
\begin{eqnarray} 
 \partial_t 
 s^{h}_{{\bf k} \parallel}
+ \Gamma_{\parallel} 
\left(
 s^{h}_{{\bf k} \parallel}
+ m^h_{{\bf k}} \right) 
=  Im \ h_{{\bf k} \parallel}(t)
+ {\bf s}^{h}_{{\bf k} \perp} \cdot \partial_t{\bf \hat{S}}
\label{s-parall-eom}
\end{eqnarray} 
and 
\begin{eqnarray} 
\left. \partial_t  {\bf s}^{h}_{{\bf k}{\perp}}\right|_{\perp}  
=\beta c {\bf S}\times{\bf s}^h_{{\bf k} \perp}
+ Im \ {\bf  h}_{{\bf k} \perp}
- \Gamma_{\perp}  
{\bf s}^{h}_{{\bf k} \perp}
- s^{h}_{{\bf k} \parallel}  \partial_t {\bf \hat{S}} 
\label{s-perp-eom} 
\end{eqnarray} 
We now focus on the 
 perpendicular component ${\bf s}^{h}_{{\perp}}$.
We note from Eq.(\ref{Mn-spin}) 
that
the motion of ${\bf S}(t)$
is characterized by a precession frequency $\omega$ determined  by 
the  field ${\bf H}$ and the mean 
hole spin $\beta {\bf s}^h_{\perp}$: 
\begin{equation} 
\partial {\bf \hat{ S}} = {\bf \omega} \times {\bf \hat{ S}} \ , \
{\bf \omega} = \beta {\bf s}^h_{\perp} - {\bf H} 
\end{equation} 
where 
\begin{equation} 
{\bf s}^h = \frac{1}{V} \sum_{{\bf k}} {\bf s}^h_{{\bf k}}
\label{sum-s} 
\end{equation} 
is 
the mean  hole spin. 
In III-Mn-V semiconductors, 
this precession 
is much slower than 
 the motion 
 of the hole spin
${\bf s}^h_{\perp}$, which is characterized by the 
precession energy $\beta c S$ and the relaxation rate $\Gamma_{\perp}$.
When considering the slower Mn spin dynamics,  we can 
then substitute 
$ {\bf s}^{h}_{{\perp}}$ by its steady state value in the frame of 
reference that rotates with the Mn spin and ${\bf \hat{S}}(t)$.
The rate of change of any vector ${\bf A}$ as seen by an observer
in the rotating frame, $\left. \partial_t {\bf A} \right|_{rot}$, 
is  related to the corresponding  rate of change 
in the inertial frame, $\partial_t {\bf A}$, by
\begin{equation} 
 \partial_t {\bf A} = \left. \partial_t {\bf A} \right|_{rot}
+ {\bf \omega} \times {\bf A}. \label{rf}  
\end{equation} 
We thus obtain in the case of the hole spin 
\begin{equation} 
 \partial_t {\bf s}^h_{\perp} = \left. \partial_t {\bf s}^h_{\perp}
 \right|_{rot}
- {\bf H} \times {\bf s}^h_{\perp}. 
\end{equation}
Projecting in the direction perpendicular to the Mn spin 
 using Eq.(\ref{perp}) 
we
 obtain that 
\begin{equation} 
\left. \partial_t  {\bf s}^{h}_{{\bf k}{\perp}}\right|_{\perp} 
= \left. \partial_t {\bf s}^h_{\perp}
 \right|_{rot,\perp} 
+ H_{\parallel} \frac{{\bf s}^h_{\perp} \times {\bf S}}{S},  
\end{equation}  
where we used  the vector property 
\begin{equation} 
{\bf A} \times \left( {\bf B} \times {\bf C} \right) 
= \left( {\bf A} \cdot {\bf C} \right) {\bf B} 
-\left( {\bf A} \cdot {\bf B} \right) {\bf C}. 
\label{cross} 
\end{equation} 
The rotating frame approximation 
 corresponds to 
neglecting  $\left. \partial_t  {\bf s}^{h}_{{\perp}}\right|_{rot,\perp}$.

To obtain an 
effective equation of motion for the Mn spin, 
we use Eq.(\ref{Mn-spin}) for 
${\bf s}^h_{\perp}
\times {\bf S}$ 
to eliminate the hole spin ${\bf s}^h_{\perp}$, 
defined by Eqs (\ref{sum-s}) and 
(\ref{perp}),  from
 Eq.(\ref{s-perp-eom}):
\begin{eqnarray}
{\bf s}^h_{\perp} 
= {\bf \hat{S}} \times \left( 
{\bf s}^h_{\perp} \times {\bf \hat{S}} \right) 
= \frac{{\bf \hat{S}} \times \partial_{t} {\bf \hat{S}}
+ {\bf \hat{S}} \times \left( {\bf H } \times 
{\bf \hat{S}} \right)}{\beta}. 
\end{eqnarray} 
By summing  Eq.(\ref{s-perp-eom}) over all momenta 
we thus obtain within the rotating frame approximation 
after using the definition Eq.(\ref{h-eff}) that 
\begin{eqnarray} 
&&\left( \beta c S +  \beta s^{h}_{ \parallel} +  
H_{\parallel}\right)
 \partial_t {\bf \hat{S}} = \left( \beta c S+ H_{\parallel} \right)
{\bf \hat{S}} \times {\bf H} 
\nonumber \\
&&
+ \beta
{\bf \hat{S}} \times \left( {\bf h} \times {\bf \hat{S}} \right)
+ \Gamma_{\perp}
{\bf \hat{S}} \times 
\left(  {\bf \hat{S}} \times  {\bf H } 
-  \partial_{t} {\bf \hat{S}} \right).
\label{eff-eom-1}
\end{eqnarray} 
The above nonlinear equation of motion 
is governed by the 
magnetic field ${\bf H}$, the 
effective magnetic field ${\bf h}$ 
determined by the interband e--h polarizations, 
 and the hole spin relaxation ($\Gamma_{\parallel}$) 
and dephasing ($\Gamma_{\perp}$) rates. 
Eq.(\ref{eff-eom-1}) can be  transformed into an 
effective time--dependent Landau--Gilbert--like  
equation of motion
by taking  the cross product of both sides 
with ${\bf \hat{S}}$ and using
the vector property 
Eq.(\ref{cross})
and the property 
 ${\bf S} \cdot \partial_t {\bf S}=0$: 
\begin{eqnarray} 
&&\left( \beta c S+ \beta s^{h}_{ \parallel} + 
H_{\parallel}\right)
 \partial_t {\bf \hat{S}}\times {\bf \hat{S}} = 
\left( \beta c S+ H_{\parallel} \right)
{\bf \hat{S}} \times \left(  {\bf H} \times {\bf \hat{S}}\right) \nonumber \\
&&+ \beta  {\bf h} \times {\bf \hat{S}}
+ \Gamma_{\perp} 
 {\bf \hat{S}} \times {\bf H } 
- \Gamma_{\perp}  \partial_{t} {\bf \hat{S}}. 
\end{eqnarray} 
Substituting the above expression into Eq.(\ref{eff-eom-1}) 
we  obtain 
Eq.(\ref{eom-eff}).

\section{} 
\label{parallel} 

In this appendix we derive the 
relaxation of 
the hole spin component $ s^h_{\parallel}$ 
parallel to the Mn spin ${\bf S}(t)$,  
 determined by Eq.(\ref{s-parall-eom}).
 We eliminate  ${\bf s}^h_{\perp}$
from Eq.(\ref{s-parall-eom})
by using  the property 
${\bf s}^h_{\perp} \cdot \partial_t {\bf \hat{S}}= 
{\bf H} \cdot \partial_t {\bf \hat{S}}/\beta$,
obtained after using the properties 
$({\bf S} \times \partial_t {\bf S})\cdot \partial_t {\bf S} =0$,  
${\bf S} \cdot \partial_t {\bf S} =
\partial_t S^2 =0$,
 and some algebra.
We also 
eliminate $\partial_t {\bf S}$
from 
Eq.(\ref{s-parall-eom}) by 
projecting both sides of Eq.(\ref{eom-eff}) with ${\bf H}$ 
and using the vector property
${\bf A} \cdot \left( {\bf B} \times {\bf C} \right) = 
{\bf B} \cdot \left( {\bf C} \times {\bf A} \right)$
 and the relations 
${\bf H}^2= H_{\parallel}^2 + H_{\perp}^2$ and 
${\bf h} \cdot {\bf H} = h_{\parallel} H_{\parallel} 
+ {\bf h}_{\perp} \cdot {\bf H}_{\perp}$. 
By summing both sides of 
Eq.(\ref{s-parall-eom}) over all momenta 
we finally obtain the equation of motion 
\begin{eqnarray} 
&& \partial_t 
 s^{h}_{ \parallel}
+ \Gamma_{\parallel} 
\left(
 s^{h}_{ \parallel}
+ m_h \right) 
+ \frac{\Gamma_{\perp} {\bf H}_{\perp}^2 
+  \beta {\bf H}_{\perp} \cdot {\bf h}_{\perp}
}{ 
\left( \beta c S +  \beta s^{h}_{ \parallel} +  
H_{\parallel}\right)^2 + \Gamma_{\perp}^2}  
s^{h}_{\parallel} \nonumber \\
&&=  h_{ \parallel} 
+ \frac{ \Gamma_{\perp} 
 \left({\bf H} \times {\bf h}
\right)_{\parallel} -
 \left( \beta c S +  H_{\parallel}\right) 
{\bf H}_{\perp} \cdot {\bf h}_{\perp}
}{ 
\left( \beta c S +  \beta s^{h}_{ \parallel} +  
H_{\parallel}\right)^2 + \Gamma_{\perp}^2}.  
\label{s-parallel-relax}
\end{eqnarray} 

\section{} 
\label{h-calc} 
In this appendix we derive the equation of 
motion that determines the time evolution of the 
effective magentic field pulse 
${\bf h}(t)={\cal E}(t) {\bf \hat{h}}(t)$
that governs the intial femtosecond magnetization 
re--orientation and relaxation.   
Using the equations of motion 
of the interband polarizations, Eqs.(\ref{Pup1}), (\ref{Pdown1}), 
(\ref{Pup2}), and (\ref{Pdown2}),
we obtain the equation of motion 
\begin{eqnarray} 
i \partial_t {\bf \hat{h}}_{{\bf k}} 
= \Omega_{{\bf k}} {\bf \hat{h}}_{{\bf k}} 
+ \frac{i}{2} {\bf \Delta}_{{\bf k}} \times {\bf \hat{h}}_{{\bf k}} 
+ \frac{{\bf \Delta}_{{\bf k}} p_{ {\bf k}}}{2} 
+ {\cal E}(t)  {\bf {\cal S}}_{{\bf k}},
\label{h-eom}
\end{eqnarray} 
where $\Omega_{{\bf k}}$ and 
${\bf \Delta}_{{\bf k}}$ 
were defined 
by Eqs.(\ref{Omega}) and (\ref{Delta}),
\begin{equation} 
p_{{\bf k}}(t) =  \mu^*_{+} P_{{\bf k} \uparrow}^{+}(t) 
+ \mu_{-}^* P_{{\bf k} \downarrow}^{-}(t) \label{h0} 
\end{equation} 
satisfies the equation of motion 
\begin{eqnarray} 
i \partial_t p_{{\bf k}} 
= \Omega_{{\bf k}} p_{{\bf k}} + 
\frac{{\bf \Delta}_{{\bf k}} \cdot {\bf h}_{{\bf k}}}{2} 
-{\cal E}(t) {\bf {\cal N}}_{{\bf k}},
\end{eqnarray}
 where 
\begin{eqnarray} 
{\cal N}_{{\bf k}}
=  \left( |\mu_{+}|^2 +  |\mu_{-}|^2 \right) 
\left[ 1 - \left(N^e_{{\bf k}}
+N^h_{{\bf k}}\right)/2 \right]
\nonumber \\
 - \left( |\mu_{+}|^2 - |\mu_{-}|^2 \right) 
\left(s_{{\bf k}z}^h - s_{{\bf k}z}^e \right),
\end{eqnarray} 
and 
\begin{eqnarray} 
&&{\bf {\cal S}}_{{\bf k} x}
= \left( |\mu_{+}|^2 +  |\mu_{-}|^2 \right) s_{{\bf k}x}^h 
+ i \left(  |\mu_{+}|^2 - |\mu_{-}|^2 \right) 
s_{{\bf k}y}^h \nonumber \\
&&+ 2  Re(\mu_{+} \mu_{-}^*) s^e_{{\bf k}x} 
+ 2 Im (\mu_{+} \mu_{-}^*) s^e_{{\bf k} y} \nonumber
 \\
&&{\bf {\cal S}}_{{\bf k} y}
= \left( |\mu_{+}|^2 +  |\mu_{-}|^2 \right) s_{{\bf k}y}^h 
 - i \left(  |\mu_{+}|^2 - |\mu_{-}|^2 \right) 
s_{{\bf k}x}^h
\nonumber \\
&&+2 Re(\mu_{+} \mu_{-}^*) s^e_{{\bf k}y} 
- 2 Im (\mu_{+} \mu_{-}^*) s^e_{{\bf k} x}, \nonumber
 \\
&&{\bf {\cal S}}_{{\bf k} z}
=-\left(  |\mu_{+}|^2 - |\mu_{-}|^2 \right) 
\left[ 1 - (N^h_{{\bf k}}+ 
N^e_{{\bf k}})/2 \right]
\nonumber \\
&&+\left( |\mu_{+}|^2 +  |\mu_{-}|^2 \right) \left( s_{{\bf k}z}^h 
- {\bf s}_{{\bf k}z}^e \right). \label{calS}
\end{eqnarray} 
We note from the 
above equations that, 
in the absence of spin polarization in the ground state, 
${\bf S}={\bf s}^h={\bf s}^e=0$, 
${\bf h}(t)$ points along the z--direction 
of optical field propagation.
On the other hand, 
in the presence of spin--polarized holes in the 
ground state as in ferromagnetic semiconductors, 
${\bf h}$  develops additional  components determined by ${\bf s}^h$, 
the Mn spin and exchange interaction 
${\bf \Delta_{{\bf k}}}$, 
and by the magnetic anisotropy.

\end{document}